# Experimental Progress on Layered Topological Semimetals

Junchao Ma[1,5], Ke Deng[2], Lu Zheng[3], Sanfeng Wu[4,*],Zheng Liu[3,*], Shuyun Zhou[2,5,*], Dong Sun[1,5,*]

1. International Center for Quantum Materials, School of Physics, Peking University, NO. 5 Yiheyuan Road, Beijing, 100871, China
2. State Key Laboratory of Low Dimensional Quantum Physics and Department of Physics, Tsinghua University, Beijing 100084, P.R. China.
3. Centre for Programmable Materials, School of Materials Science and Engineering, Nanyang Technological University, Singapore, 639798, Singapore.
4. Department of Physics, Massachusetts Institute of Technology (MIT), Cambridge, MA 02139, USA.
5. Collaborative Innovation Center of Quantum Matter, Beijing, China.

**Abstract**

We review recent experimental progresses on layered topological materials, mainly focusing on transitional metal dichalcogenides with various lattice types including 1T, $T_d$ and 1T' structural phases. Their electronic quantum states are interestingly rich, and many appear to be topological nontrivial, such as Dirac/Weyl semimetallic phase in multilayers and quantum spin hall insulator phase in monolayers. The content covers recent major advances from material synthesis, basic characterizations, angle-resolved photoemission spectroscopy measurements, transport and optical responses. Following those, we outlook the exciting future possibilities enabled by the marriage of topological physics and two dimensional van der Waals layered heterostructures.

## 1. Introduction

Two dimensional (2D) materials and topological materials are two important areas attracting enormous research interest in materials sciences and condensed matter physics [1-5]. Although the overlap of these two fields dates back to layered topological insulator e.g. $Bi_2Se_3$ [6, 7], it's not until the recent experimental verification of type-II Dirac/Weyl semimetal and quantum spin hall insulator in 2D layered transitional metal dichalcogenides (TMDCs) family that these two fields have merged together unprecedentedly [8-11]. The emergence of 2D layered topological materials provides excellent opportunities to explore and engineer topological properties of quantum materials, as 2D layered materials can be conveniently integrated into van der Waals heterostructures by vertically stacking [12-14]. Such artificial van der Waals solids may lead to high performance functional quantum devices and enable many interesting applications of quantum phenomena. The physical properties of Dirac/Weyl semimetals have been discussed in a few comprehensive reviews [15-18]. In this review, we focus on particularly the experimental progress on Dirac/Weyl semimetals of 2D layered material candidates, *e.g.* semimetallic transitional metal dichalcogenides. We cover a wide range of experimental aspects, including synthesis, basic material characterization, unique topological band structure and surface states as revealed by angle resolved photo emission spectroscopy (ARPES) measurement, and interesting quantum and topological phenomena revealed by transport and optical measurements. As the experimental investigation regarding Dirac/Weyl semimetals is still at its early stage, a large amount of effort focuses on experimental testing and verification of the Dirac/Weyl semimetal candidates, and many other interesting topics remain to be explored. Special attention will be drawn on their optical response in this review, although related experimental reports are relatively rare. In light of this, we will also include some discussions on representative experimental work on bulk topological semimetals, especially in the optical response part, from which we hope to provide a guidance to the study of the 2D layered counterparts. We note that there is a parallel theoretical review on relevant topics by Qian *et al*, hence theoretical aspects are not at the center of gravity here. The goal of this review is to summarize the existing experimental progress with a hope to shine light for future developments. Several future experimental possibilities are discussed at the end of this review.

## 2. Concept of Topological Semimetals

In Landau's classification, states of matter are classified through the principle of spontaneous symmetry breaking, e.g. ferromagnetic states are related to time-reversal symmetry breaking. After the discovery of Quantum Hall Effect (QHE) [19], it has been recognized over the past decades that such classification is incomplete, and the concept of topology has to be introduced to understand condensed matter systems. For example, it is necessary to classify two-dimensional (2D) electronic systems by their topological invariants determined from their band structures (e.g. Chern number or TKNN number) [20, 21]. In addition, the marriage between topology and symmetry leads to the understanding of a large family of materials with novel physics. An excellent example is time reversal invariant topological insulators (TI), a peculiar type of insulators with nontrivial band topology. Experimentally, 2D TI exhibits quantum spin Hall effect (QSHE), while three dimensional (3D) topological insulators feature helical surface mode [22, 23]. Such topological insulators are characterized by a topological invariant called $\boldsymbol{Z_2}$ index and topological protected gapless surface states [22, 23]. Further studies revealed that the topological classification of the band structure can be extended to semimetals, leading to the concept of Dirac semimetals, in which 3D Dirac fermions emerge [24-30]. The Dirac fermions in 3D Dirac semimetals are protected by the crystal symmetry of the bulk crystal. If inversion- or time-reversal symmetry is broken, each Dirac fermion can be split into two Weyl fermions. Weyl fermions have been realized in TaAs family [18, 31, 32]. While 3D Dirac semimetal states exist if the related symmorphic symmetry is preserved [18, 33], as critical points of topological phase transitions, Weyl semimetal states are expected to be more robust since their gapless band crossings are pairs of topological defects, and they cannot be gapped out unless they are annihilated in pairs [24]. More recently, it was realized that topological semimetal can be further categorized into two types according to Lorentz invariant [15, 16, 34]. In type-I Dirac/Weyl semimetal states, the linear Dirac cone obeys Lorentz invariant, while in type-II Dirac/Weyl semimetal, the Dirac cone is significantly tilted and cannot be adiabatically transformed to the untilted one, thus breaking the Lorentz invariance[11, 33, 35-39]. In addition, both type-II Dirac and Weyl fermions exist in solids and have no counterparts in high energy physics, thus providing a model system for investigating new topological states beyond the standard model.

TMDCs with the chemical formula of $MX_2$ (M=Mo, W; X=S, Se and Te) have attracted extensive research interests with intriguing properties for electronics, optoelectronics and valleytronics applications in the past decade [4, 40-46]. It has been realized that strong spin-orbit coupling of some TMDCs, together with the rich crystal structures, e.g. hexagonal (H), trigonal(1T), distorted trigonal (1T') and tetragonal ($T_d$) structure, can provide an interesting platform for realizing new topological phases and beyond. Under ambient conditions, $MoS_2$ and $WSe_2$ favor 2H phases which are often semiconductors [47]; $TaS_2$ can possibly exist in several stable forms including 2H and 1T, whose phase diagrams harbor a number of interesting states ranging from charge density wave [48], superconductivity to even possibly quantum spin liquid [49]; $MoTe_2$ favors semiconducting 2H phase or metallic 1T' phase (which undergoes a transition to $T_d$ under low temperature) [10]; and $WTe_2$ is a semimetal stabilized at $T_d$ phase. One excellent example of their topological quantum phases can be found in 1T'-$MX_2$ monolayers, which have been proposed to be a large gap 2D topological insulator that hosts quantum spin Hall effect back to 2014 [8]. The experimental study of such systems has been achieved very recently [50-55]. In the meantime, the investigation of type-II 3D Dirac/Weyl fermions in the bulk 1T' and $T_d$ structured $MX_2$ have developed rapidly during the past few years [9-11, 16, 18, 34, 36, 56]. In this review, we focus on these recently discovered topological states in TMDCs, including 3D Weyl fermions, 3D Dirac fermions and 2D topological insulators, as summarized in Table I below.

**Table 1.** A summary of topological states realized in transition metal dichalcogenides: 3D Dirac semimetal, 3D Weyl semimetal and 2D topological insulator (top row), the host materials (middle row) and the corresponding crystal structure (bottom row).

| 3D Dirac semimetal | 3D Weyl semimetal | 2D Topological insulator (QSHI) |
|---|---|---|
| 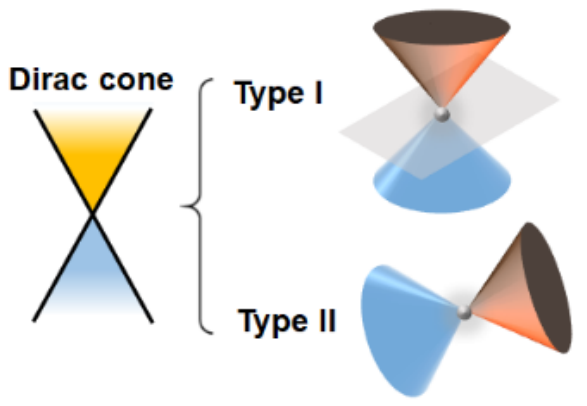 <br>Dirac fermions:<br>Protected by crystal symmetry | 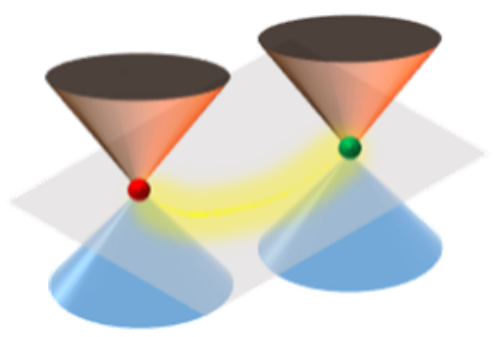<br>Weyl fermions:<br>Broken time- or inversion-symmetry | 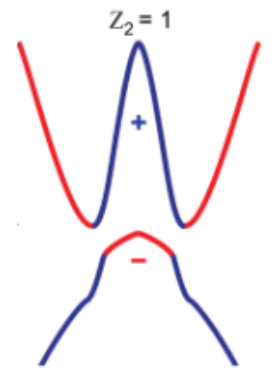 <br>2D TI:<br>Band inversion + SOC |
| **Type II Dirac semimetals (bulk):**<br>• 1T-$PtTe_2$<br>• 1T-$PtSe_2$<br>• 1T-$PdTe_2$ | **Type II Weyl semimetals (bulk):**<br>• $T_d$-$MoTe_2$<br>• $T_d$-$WTe_2$<br>• $T_d$-$W_{1-x}Mo_xTe_2$ | **2D TI (film):**<br>• 1T' ($T_d$)-$WTe_2$<br>• 1T'-$WSe_2$<br>• 1T'-$MoTe_2$ |
| 1T 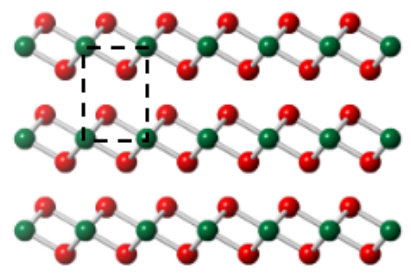 | $T_d$ 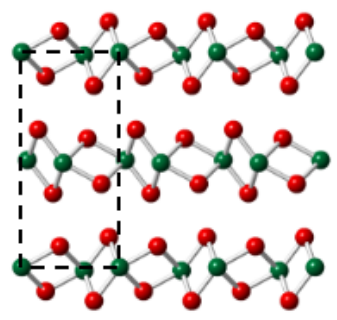 | 1T' 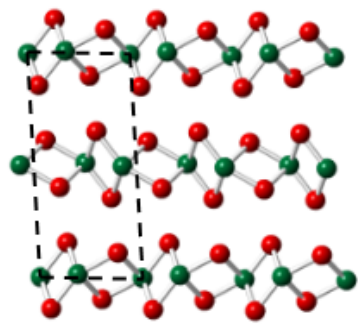 |

## 3. Synthesis methods of basic material characterization

Currently, a few synthetic strategies have been applied to prepare 2D Layered materials, such as solid-state reaction, molecular beam epitaxy (MBE), chemical vapor deposition (CVD) and chemical synthesis. 2D TMDCs typically crystallize into 2H, 1T, 1T' and $T_d$ phases. The 2H and 1T phase are primarily semiconducting as the most common structure, while the 1T' and $T_d$ compounds are typically semimetallic with the salient quantum phenomena [57]. Among TMDCs, $MoTe_2$ and $WTe_2$ have been widely studied in terms of edge states, magneto-transport and phase transition.

The solid-state reactions, including physical vapor transport (PVT) [58-60] and chemical vapor transport (CVT) [61-64], are the most widely used and effective methods to grow single crystal 2D materials with the assistance of small amount transport agency (figure 1(a)). High-quality layered 2D Dirac/Weyl semimetals like $WTe_2$ [65], $MoTe_2$ [57], $PtSe_2$ [11], $PtTe_2$ [33], and even ternary TMDC alloys such as $TaIrTe_4$ [39, 66], $W_xMo_{1-x}Te_2$ [67] have been obtained by solid-state reaction. Most Mo and W-dichalcogenides are stable in the semiconducting 2H phase. $WTe_2$ tends to crystalize into $T_d$ phase. Both 2H and 1T' phases can be found in $MoTe_2$ and both are thermodynamically stable. 2H-$MoTe_2$ and 1T'-$MoTe_2$ can be synthesized by flux based solid-state reaction. During the synthesis of $MoTe_2$, a low cooling rate results in 2H-$MoTe_2$, while the rapid quench cooling yielding the 1T'-$MoTe_2$. Monolayer and few layer TMDCs can be achieved by mechanical exfoliation and solution exfoliation (figures 1(b) and (c)). Mechanical exfoliation with “scotch tape” was first developed to exfoliate single layer graphene [68], and now widely used in cleaving bulk layered materials to obtain high quality samples [69-71]. Since monolayer and bilayer $WTe_2$ degraded quickly in air [72], the exfoliation is generally carried out in glove box under Ar atmosphere [73]. Yu *et al* presented an acetone exfoliation of bulk single crystal $T_d$-$WTe_2$ into nanosheets [63].

Complimentary to solid reactions, MBE and CVD are promising techniques for large-scale and high-quality 2D materials and van der Waals heterostructures. MBE technique attracted increasing attention due to the epitaxial thickness and doping control for desirable large and high-quality samples, such as $MoS_2$ [74], $MoSe_2$ [75], $WSe_2$ [76], and $MoTe_2$ [77]. More importantly, MBE can realize various heterostructures with pre-designed 2D blocks [78, 79]. However, the challenges in the MBE growth lies on that the different vapor pressure between different source materials and the low melting points of sulfur precursors, which results into a narrow “growth window” and strict growth conditions. Therefore, only a few TMDCs can be grown by this method. The high-quality single-crystal monolayer $PtSe_2$ (Type-II Dirac Semimetal) has been obtained by direct selenization of Pt substrate at 270 ℃ [80] and MBE [81]. $PtSe_2$ [82] and $PtTe_2$ [83] can also be synthesized by CVD method [84-90]. By adjusting experimental parameters, *e.g.* gas flowing rate, pressure, growth temperature and growth time, various 2D ultrathin layers can be achieved (figure 1(d)). Recently, Zhou *et al* demonstrated that the molten-salt-assisted CVD method could be broadly applied for the synthesis of 2D atomically thin transition-metal chalcogenides libraries ranging from semiconductors, semimetals, metals to superconductors [91]. They synthesized 47 TMDC compounds including 32 binary compounds (based on the transition metals Ti, Zr, Hf, V, Nb, Ta, Mo, W, Re, Pt, Pd and Fe), 13 alloys (including 11 ternary, one quaternary and one quinary), and 2 heterostructure compounds. Monolayer 2H and 1T'-$MoTe_2$ and $T_d$-$WTe_2$ are synthesized by CVD [92], and 2H and 1T'-$MoTe_2$ can be controlling synthesized by controlling the flowing rate of Te source [57].

Apart from the methods above, hydrothermal method is also proved to be a facile and low-cost approach to synthesize 2D materials, such as $MoS_2$ [93], $MoSe_2$ [93], $WS_2$ [94], and $WSe_2$ [95], 1T'-$MoTe_2$ [96] and $WTe_2$ [97]. Being an alternative way, hydrothermal method can produce a large amount of high-quality TMDC samples under low reaction temperature.

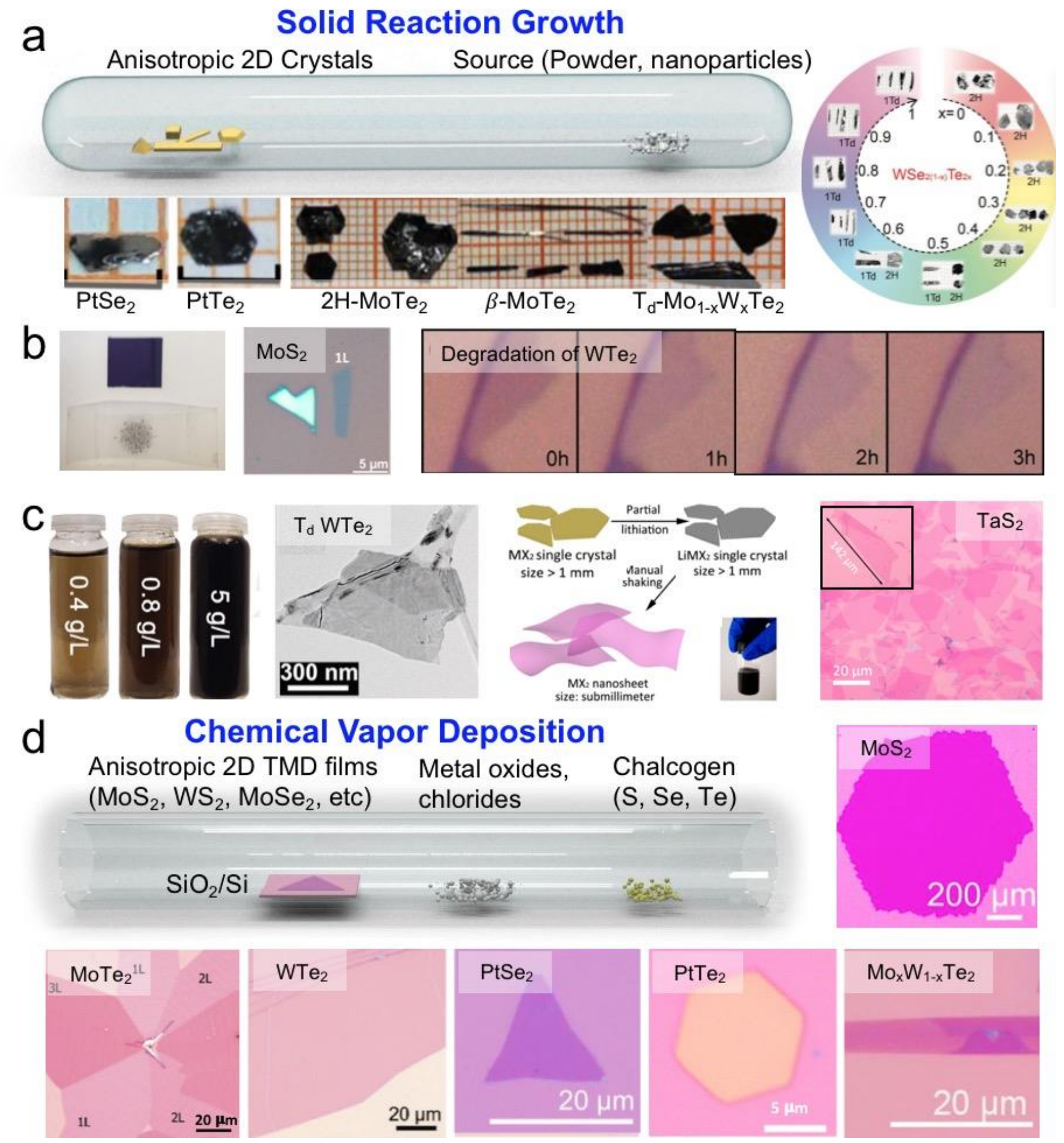

**Figure 1.** (a) Schematic of single crystal growth of TMDC by PVT/ CVT method, and the images of single crystals of $WSe_{2(1-x)}Te_{2x}$ [62], © 2016 John Wiley and Sons, $PtSe_2$ [11], $PtTe_2$ [11], © 2017 American Physical Society, 2H-$MoTe_2$ [67], $\beta$-$MoTe_2$ [67], $T_d$-$Mo_{1-x}W_xTe_2$ [67], © 2017 Nature Publishing Group. (b) Monolayer $MoS_2$ obtained by scotch tape mechanical exfoliation [71], © 2014 American Chemical Society, and degradation of exfoliated $WTe_2$ [71], © The Royal Society of Chemistry 2016. (c) The $WTe_2$ nanosheets with different concentrations by ultrasonic exfoliation and the corresponding TEM image [63], © 2017 John Wiley and Sons, and $TaS_2$ nanosheets obtained by partial lithiation reaction [70], © 2017 American Chemical Society. (d) Schematic of single crystals growth by CVD method, and the obtained $MoSe_2$ [91], © 2018 Nature Publishing Group, $MoTe_2$ [92], $WTe_2$ [92], © 2016 John Wiley and Sons, $PtSe_2$ [91], $PtTe_2$ [83], © 2018 American Chemical Society, $Mo_xW_{1-x}Te_2$ [91].

# 4. Basic Material Characterizations

Characterization techniques such as optical microscopy (OM), Raman scattering spectroscopy, photoluminescence (PL), atomic force microscopy (AFM), transmission-electron microscopy (TEM) and scanning transmission electron microscope (STEM) are widely employed to characterize 2D materials, as shown in figure 2.

Optical microscopy provides a simple, swift and reliable way to identify the geometries and estimate the thicknesses of 2D materials like graphene, $MoS_2$, $WSe_2$ and $TaS_2$. [84, 85]. Based on the contrast, one can estimate the layer numbers [98, 99]. As shown in figure 2(a), the optical microscopy image of an ultrathin $MoS_2$ flake from 1-15 layers on 90 nm $SiO_2$/Si are presented. The accurate thickness can be determined by AFM height topography.

Being a powerful and non-destructive technology, Raman spectroscopy is usually employed to identify the species of 2D materials, the quality of 2D materials and, more interestingly, the inversion symmetry breaking in layered Weyl semimetals. Raman spectra contains the information of not only structural properties, but also the layer thickness, band structures, strain effects, doping type, concentration, electron-phonon coupling, and interlayer coupling [100]. Fu *et al* systemically studied the thickness-dependent charge density wave (CDW) phase transitions of 1T-$TaS_2$ by temperature-dependent Raman spectra and plot the thickness-temperature phase diagram [101]. Raman spectra are sensitive to crystal symmetry, therefore it can be useful to reveal the inversion symmetry breaking, considering that the noncentrosymmetry is a prerequisite for non-magnetic Weyl semimetals [102]. Zhang *et al* presented direct spectroscopic evidence for the inversion symmetry breaking in the low-temperature phase of $MoTe_2$ by systematic Raman experiments [103]. Additionally, the photoluminescence (PL) spectroscopy is commonly used to evaluate the band structure, impurity/defect impacts of materials [4, 86, 104]. For example, the strong PL can be obtained from direct band gap semiconductors, while it is difficult for indirect semiconductors. This phenomenon was first observed by Wang *et al* and Mak *et al* in $MoS_2$ [105, 106], and following many monolayer semiconductors show the similar properties including $MoSe_2$ [75, 107], $WS_2$ [108-110], and $WSe_2$ [111, 112].

AFM are mostly employed to identify thickness of 2D materials [113]. It also can be used to determine the surface potential, modulus, adhesion, conductivity, IR absorption and reflection of 2D materials [85, 86, 88]. With the special tips, AFM even works under liquids as well as in the ambient pressure. Interestingly, Wang *et al* studied the surface potential and vertical piezoelectricity of CdS thin films by scanning Kelvin force microscopy (SKFM) and piezoelectric force microscopy (PFM) and demonstrated the vertical piezoelectric domains at the CdS thin films with piezoelectric coefficient ($d_{33}$) was 32.8 pm $\cdot V^{-1}$, which value is nearly three times larger than that of the bulk one [114].

TEM (transmission electron microscopy) and STEM (scanning transmission electron microscopy) are often used to evaluate the atomic structure and composition of 2D materials, such as crystal structure, surface properties, interlayer stacking relationships, domain sizes, and elemental configuration [115-118]. In TEM, selected area electron diffraction (SAED) patterns can resolve the crystal structures of 2D materials, as well as the crystallographic orientation between two crystals [62, 116, 119]. Electron energy loss spectroscopy (EELS) and Energy dispersive X-ray spectroscopy (EDX) are used to image an individual or separate atom in a single layer [63, 91, 120, 121]. Due to its very high lateral resolution and low accelerating voltage, STEM becomes more popular to obtain atomically resolved images of 2D materials. Atomic arrangements of the 2H- and 1T'-$MoTe_2$ can be seen by STEM, as shown in figure 2(d), in which the unit cells are clearly related to the simulated unit cells [57]. Also, $MoS_2/WSe_2$ van der Waals hetero-bilayer structures with different local rotational alignment can be resolved by STEM, which exhibited the Moiré patterns with well-defined periodicity simulations [118].

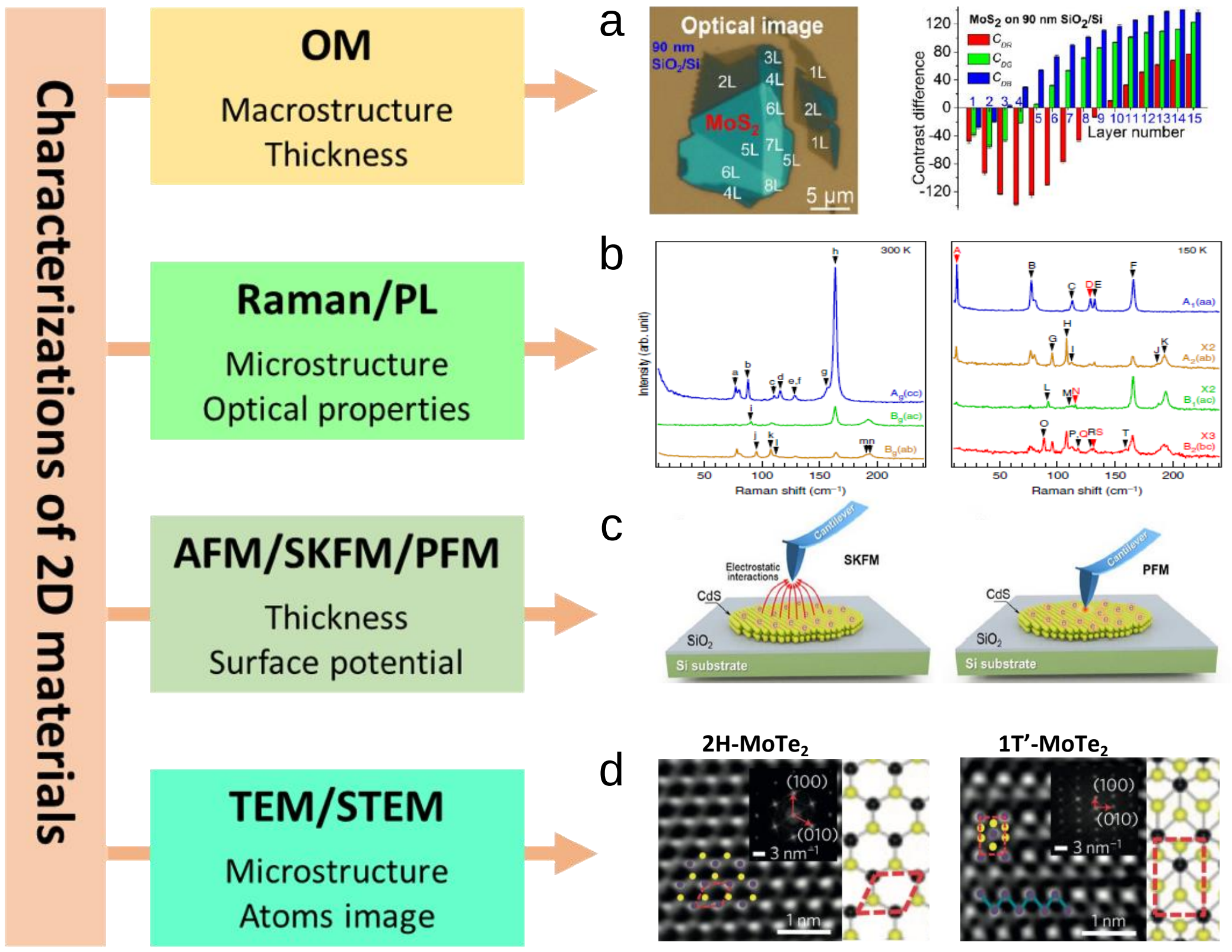

**Figure 2.** (a) Thickness identification of $MoS_2$ by optical image [99], © 2013 American Chemical Society. (b) Polarized Raman spectra measured in the high- and low-temperature phases of $MoTe_2$. The red labels A, D, N, Q and S mark the Raman modes that are directly sensitive to the inversion symmetry breaking [103], © 2016 Nature Publishing Group. (c) Schematic of SKFM and PFM measurements for ultra-thin materials [114], with permission from AAAS. (d) TEM bright-field images of 2H-$MoTe_2$, 1T'-$MoTe_2$ and their related simulated unit cell [57], © 2015 Nature Publishing Group.

## 5. ARPES verification and characterization of Band structure and Surface Sates

Angle-resolved photoemission spectroscopy (ARPES) is a powerful tool for resolving the electronic structure of both the bulk and surface states, and a comparison of ARPES results with theoretical calculation has been critical for verifying these intriguing topological phases, e.g. Dirac/Weyl semimetals. Therefore, in this section, we will focus mainly on the ARPES evidence for experimental realization of these topological phases. In a 3D topological Dirac semimetal, the 3D Dirac band touching arises from certain symmetry protection, giving rise to linearly dispersing bulk band, with double surface states connecting a pair of Dirac points forming closed Fermi surface. The 3D Dirac semimetal state has been confirmed in $Na_3Bi$ and $Cd_3As_2$ systems by observing the linear band dispersion and closed double Fermi arcs [29, 122-124]. Moreover, 3D Weyl semimetals possess pairs of linear dispersing bulk band structure and open topological surface states. ARPES shows its advantage in distinguishing those features which has led to the observation of Weyl semimetallic phase in TaAs. The observation of Fermi arc features and pairs of linearly dispersed bulk band serve as the smoking gun evidences of Weyl semimetal states [31, 32, 125].

### 5.1. Type-II Weyl semimetal

In a Weyl semimetal, Weyl points carry a left- or right-handed chirality and they always appear in pairs of opposite chirality according to a no-go theorem [126]. One intriguing property of Weyl semimetal is that it has open Fermi surface called Fermi arc, which is formed by topological surface state [3] and connects the projected Weyl points with different chiralities [3]. Following the realization of Weyl fermions, it was realized theoretically that Weyl fermions can have two different types [34]. Different to the type-I Weyl fermion observed in TaAs, type-II Weyl fermions has a strongly titled Dirac cone along certain momentum direction (figure 3) and thereby breaking the Lorentz invariance. This new type of Weyl fermion emerges at the touching points between electron and hole pockets, and therefore there is finite density of states at the Weyl point.

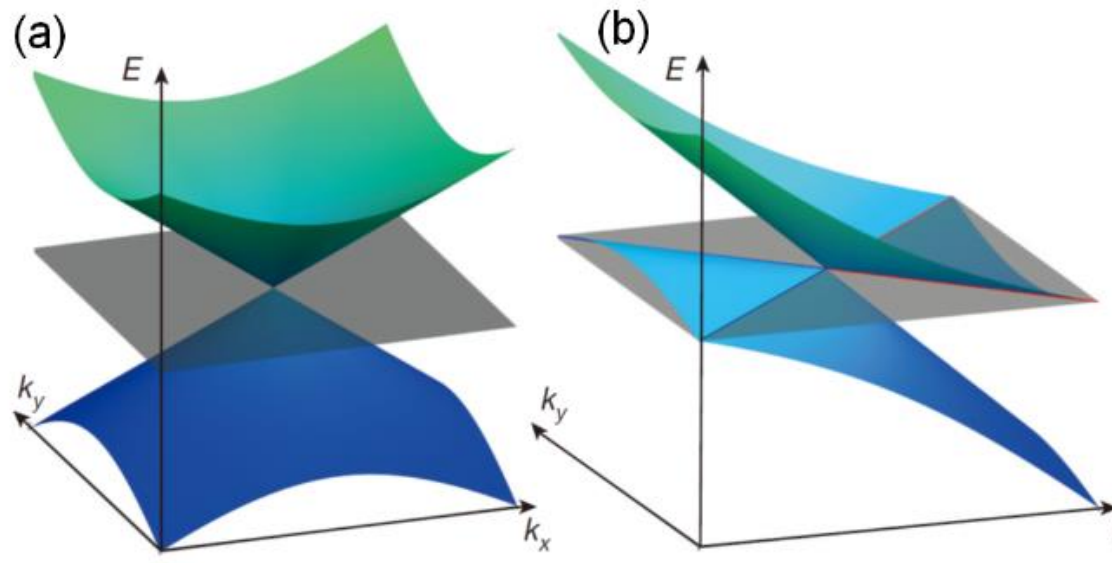

**Figure 3.** (a) Illustration of type-I electronic band dispersion, (b) type-II electronic band dispersion. Figures (a) and (b) are reproduced with permission from [34], © 2015 Nature Publishing Group.

The existence of type II Weyl fermion was first proposed in $WTe_2$ [34]. However, the separation of the proposed Weyl points in the momentum space is very small (0.7% of the reciprocal lattice vector), making it extremely challenging to resolve the Fermi arc feature via ARPES measurement [56, 127]. Luckily, theoretical calculation suggests that the Fermi arc features of another candidate, $T_d$-$MoTe_2$, are much larger and potentially accessible for spectroscopic studies [56, 127]. $MoTe_2$ is usually stable at 1T' phase at room temperature, but it undergoes a structural phase transition at ~ 250 K from 1T' to possibly $T_d$ phase (similar to $WTe_2$) [128] that can break the inversion symmetry. The symmetry breaking has been confirmed by observing the emergence of infrared-active Raman modes below the structural phase transition temperature [103], which is verified in a number of Raman spectroscopic studies [67, 102, 129-134].

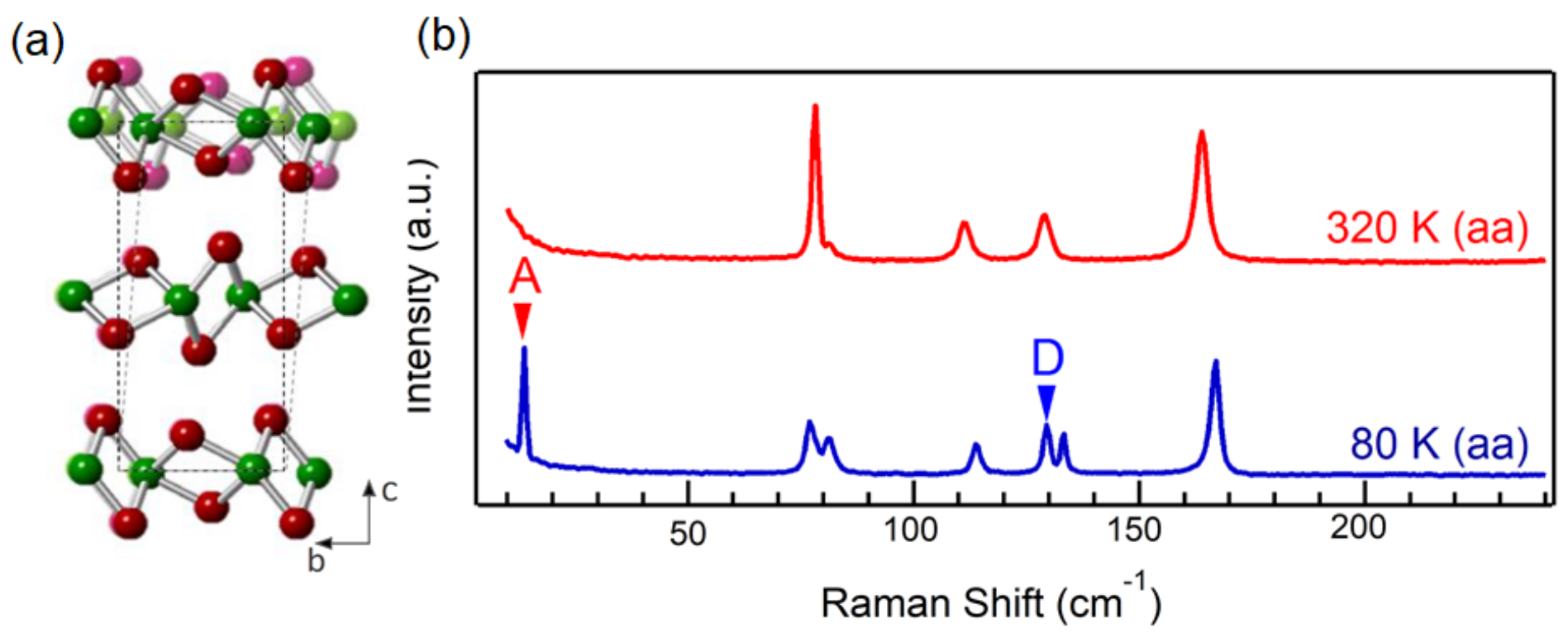

**Figure 4.** (a) Illustration of structural phase transition of $MoTe_2$. (b) Raman spectra at 320 K and 80 K respectively, A and D marked with arrows indicate the Raman-inactive mode in 1T' phase emerge at low temperature. Figures (a) and (b) are reproduced with permission from [103], © 2016 Nature Publishing Group.

The most challenging part in revealing type-II Weyl semimetal is that the Fermi arcs emerge in a small region between the bulk electron and hole pockets [56, 127]. To identify the topological surface states, several surface-sensitive techniques are employed. Signatures of the Fermi arcs and topological surface states are identified in ARPES (figure 5) by using different photon energies. In particular, laser source at 6.2 eV is sensitive to the bulk states while ultraviolet photon energy from synchrotron are surface sensitive and critical for revealing the surface states [10]. A comparison of the calculated band structures for the 1T' and $T_d$ phases allows one to distinguish the trivial surface states from the topological Fermi arcs. The topological states are expected to exist only in the low temperature phase and are connected to the Weyl points. The experimentally measured termination points and the evolution as a function of energy match well with the calculation (figure 5(c)) [10]. Other ARPES works also show results that are consistent with the topological surface states and Fermi arc [38, 135-140]. Additional supporting evidence is provided by utilizing quasiparticle interference (QPI) obtained via scanning tunneling spectroscopy. Scattering vectors $\mathbf{q}_1$ and $\mathbf{q}_2$ which connect the termination points of the Fermi arcs (figures 5(d) and (e)) are identified and their dispersions suggest that they are scattered from a pair of topological Fermi arcs [10]. More extensive QPI results can be found in later works [141-143]. The signatures of Fermi arcs from two complementary surface-sensitive techniques (ARPES and STM), and their good agreement with the calculated results, establish $T_d$-$MoTe_2$ as a type-II Weyl semimetal [10].

Since $MoTe_2$ in the $T_d$ phase breaks the inversion symmetry, different terminated surfaces (0 0 1) and (0 0 -1) will give rise to different polarities and very different connectivity of Weyl points. Experimental evidence for the different surface termination has been revealed by several groups [136, 144]. In addition to the establishment of type-II Weyl semimetal state in $T_d$-$MoTe_2$, many efforts have also been made in the investigation of $WTe_2$. Although still under debate, topological surface states together with trivial surface state have been revealed by ARPES and STM [9, 145-150]. The doping evolution of $WTe_2$ and $MoTe_2$ has also been investigated. By doping $WTe_2$ with

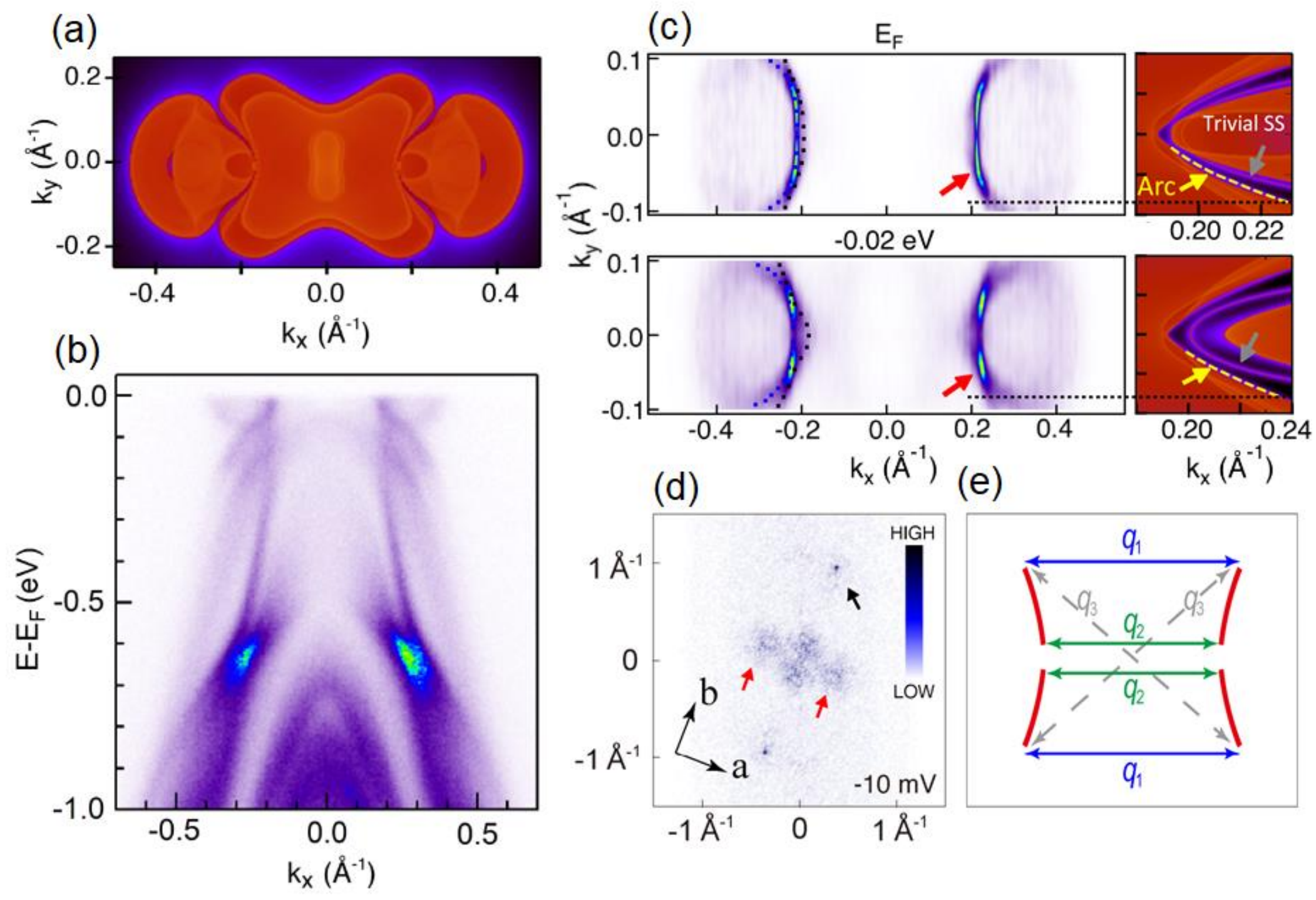

**Figure 5.** (a) A calculated Fermi surface of $T_d$-$MoTe_2$. (b) ARPES measurement of a band structure along $\Gamma - X$ direction, (c) The constant energy map with comparison with calculation results. Arrows point out the nontrivial Fermi arc states. (d) The FFT power spectra of the d$I$/d$V$ maps from STS of $T_d$-$MoTe_2$, and shows the QPI pattern of $T_d$-$MoTe_2$. (e) An illustrative figure to show all the extremal pairs owing to a pair of topological Fermi arcs. Figures (a)-(e) are reproduced with permission from [10], © 2016 Nature Publishing Group.

Mo, the Fermi arcs can be tuned to be larger in comparison to $WTe_2$, assuming that the crystal structure of $WTe_2$ is preserved upon doping [35, 151]. Recent systematic characterization of W-doped $MoTe_2$ suggests that the $T_d$ phase can be stabilized at room temperature with 8% W substitution [152]. The doping of $MoTe_2$ by Nb has also been investigated and the phase diagram has been obtained [153].

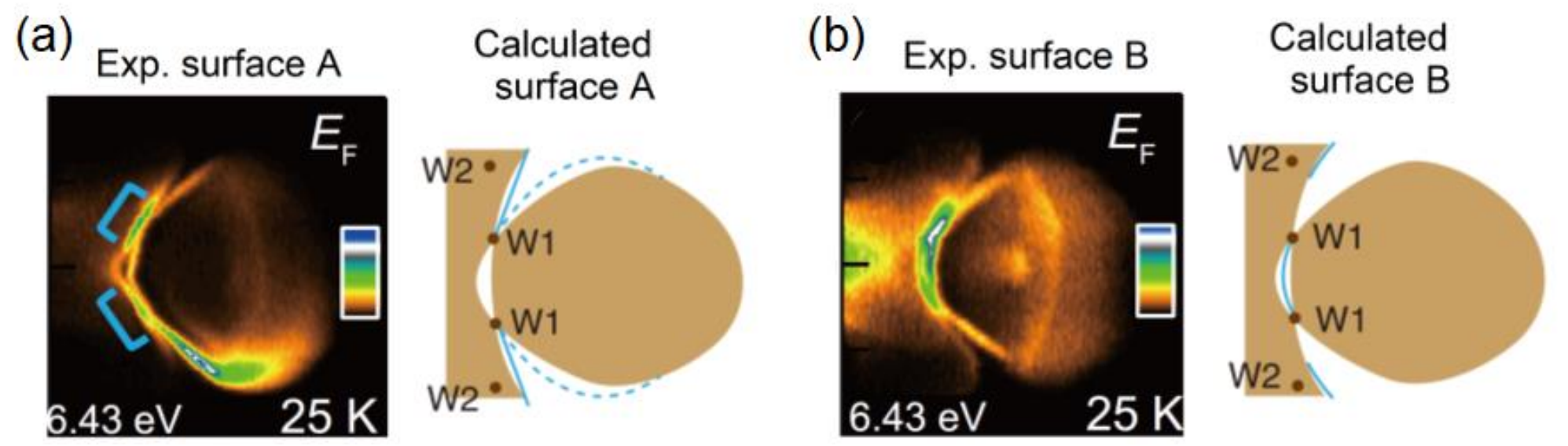

**Figure 6.** (a), (b) different Fermi surface map on a specified range to show different connectivity and arc configuration in (0 0 1) and (0 0 -1) surface, respectively. Figures (a) and (b) are reproduced from [144], © 2017 American Physical Society.

### 5.2. Type-II Dirac semimetal

The classification of type-II versus type-I can also be extended to Dirac semimetals. Soon after the discovery of type-II Weyl semimetals in $MoTe_2$, type-II 3D Dirac fermions were reported in $PtTe_2$ by Zhou group [33]. ARPES measurements reveal two Dirac cones in the 3D momentum space at about -1 eV below $E_F$. These two Dirac cones emerge at the intersection between the electron and hole pocket and they are strongly tilted along the $\mathbf{k}_z$ direction. Being similar to $PtTe_2$, $PtSe_2$ has also been established as a type-II Dirac semimetal both theoretically and experimentally [11, 36]. Another cousin of $PtTe_2$ and $PtSe_2$, $PdTe_2$ not only has the same crystal symmetry and resembled band structure [154, 155], but also is realized later to have nontrivial Berry phase [155-161]. Thus it is also believed to be a type-II Dirac semimetal. Intriguingly, if reducing the thickness of the 3D Dirac semimetal $PtSe_2$, it will go through a crossover from Dirac semimetal in 3D to a semiconductor at the 2D limit [80, 81, 162-164]. In the single layer condition, helical spin texture with spin-layer locking induced by local Rashba (R2) effect has also been revealed in monolayer of $PtSe_2$ [81, 165].

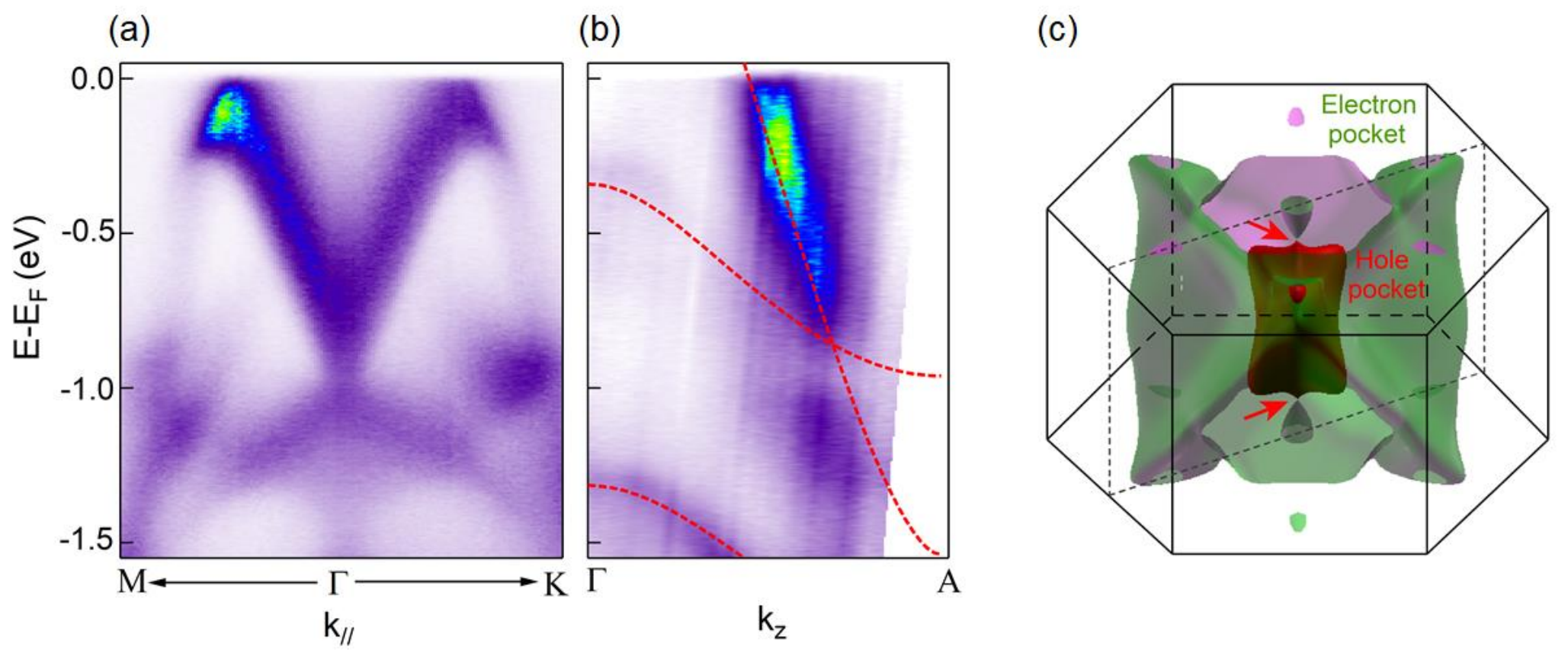

**Figure 7.** (a) In plane Dirac cone along $M - \Gamma - K$ direction. (b) Measured out of plane dispersion along $\Gamma - A$ direction with $\mathbf{k}_{//}$=0, shows a tilted band crossing. Red dashed lines are calculated dispersion along $\Gamma - A$ direction. (c) Three-dimensional plot of the electron and hole pockets at the Dirac point energy. The electron and hole pockets touch at the Dirac point. Figures (a)-(c) are reproduced from [33], © 2017 Nature Publishing Group.

### 5.3. Monolayer Quantum Spin Hall insulator

Before the realization of 3D Weyl and Dirac semimetals, quantum spin hall effect (QSH) has already been predicted in single layer distorted trigonal structure 1T'-$MX_2$ (M=W and Mo, X=S, Se and Te) in 2014 [8]. Unlike hexagonal semimetallic graphene, monolayer 1T'-$MX_2$ is monoclinic and expected to be a 2D topological insulator (TI) due to the band inversion caused by the distorted structure and a sizeable gap opening is expected near the Fermi level (thanks to the strong spin-orbital coupling of metal *d* orbital). A helical surface state has been predicted inside in the gap [8, 50]. In addition, the inverted band and nontrivial topology are predicted to be tunable by strains, electric fields etc. Such tunability might enable the application in field effect transistor (FET) [8, 50]. Most tantalizingly, Van der Waals heterostructure formed by 1T'-$MX_2$ is proposed to be topological field effect transistors (vdW-TFET) [8].

Recently, ARPES measurement of this novel 2D quantum state has been realized in the epitaxy grown monolayer $WTe_2$ film [51]. The measurement (figure 8(b)) confirms the predicted signature of topological band inversion (figure 8(a)), and a gap of 45 meV has been revealed (figure 8(c)). Scanning tunneling microscopy (STM) measurements further confirm the insulating bulk state with a gap opening and the metallic edge state (figure 8(d)), providing additional support for the 2D TI or QSHI [51]. The recent quantum transport measurements provide another direct way to study the 2D topological state, which we summarize in the next section. Not only monolayer $WTe_2$ can achieve QSHE, but also other compounds with same symmetry and proper spin-orbit coupling. Monolayer $T_d$-$WTe_2$ film serves as the first established 2D TI in TMDC compounds, and its cousins are expected to behave similar [53, 55, 166, 167], although further experimental evidence is needed for verification.

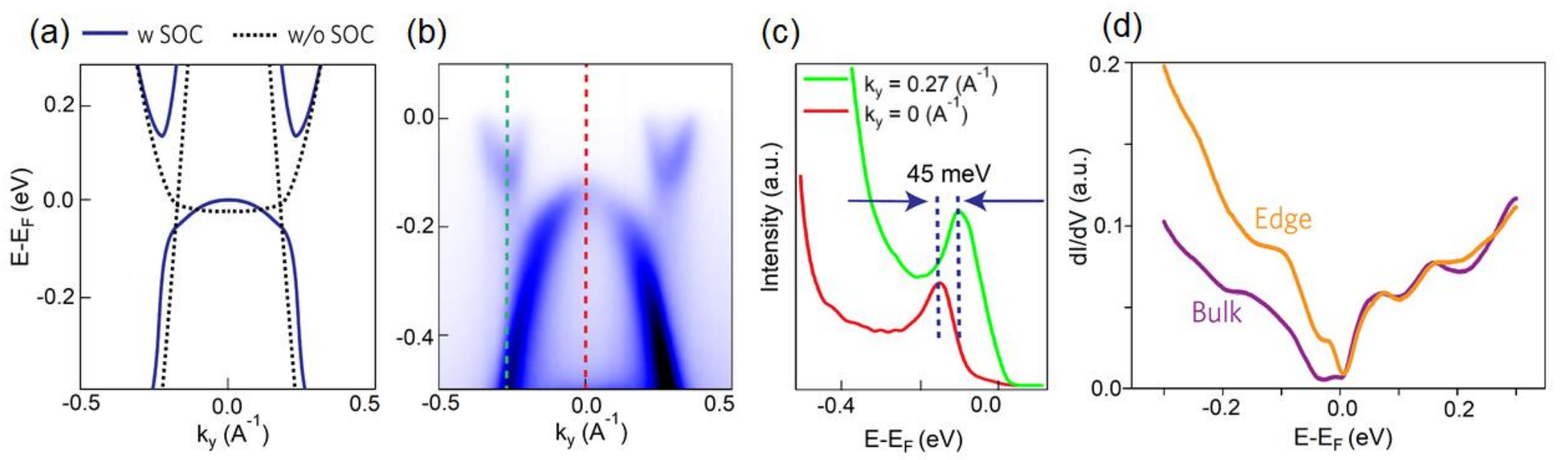

**Figure 8.** (a) calculated band structure along $\Gamma - Y$ direction of monolayer 1T'-$WTe_2$. (b) ARPES measured band structure along $\Gamma - Y$ direction of monolayer 1T'-$WTe_2$. (c) Energy distribution curves (EDCs) from the momentum positions marked with green and red lines in (b). (d) Representative dI/dV spectra taken at the edge (orange) and in the bulk (purple), respectively. Figures (a)-(d) are reproduced with permission from [51], © 2017 Nature Publishing Group.

## 6. Transport Properties

In this section we review the recent quantum transport measurements of layered topological materials, with emphasis on TMDCs. The nontrivial topology of electron wavefunctions in topological materials can give rise to highly specific, unusual transport behaviors that are protected from local perturbations. The remarkable examples are the precisely quantized Hall resistance in integer and fractional quantum Hall effects and quantum anomalous Hall effect. These rare examples, however, have uncovered the possibilities of a zoo of rich topological quantum states [168], even at zero magnetic field. One interesting situation is at the intersection of topology and correlations. This is appealing because excitations in correlated many-body systems can behave as a fraction of one electron [169], although a free electron is fundamentally undividable. Fractionalized particles in topological materials, such as Majorana modes in topological superconductors, can in principle provide a way to encode quantum information nonlocally, immune to errors [170]. To date, it remains to be an outstanding goal to identify many novel topological states, especially those with fractionalized excitations.

A promising direction is to look into 2D layered materials. Electron correlations have known to be essential in many layered materials, among which the most well-known examples are perhaps cuprates and iron-based superconductors. Theoretically, there also exists a large number of predicted 2D materials with nontrivial topological properties [8, 171-178]. In fact, it is the prediction of identifying graphene as a quantum spin Hall insulator (QSHI) in 2005 [171], together with others [22, 23, 179], that gives rise to the notion of topological insulators. However, experimental study of many predicted 2D layered topological materials remains difficult, often due to their unstable structures. Recent experimental and theoretical progresses on $T_d$ or 1T' layered TMDCs, in particular $WTe_2$ and $MoTe_2$, have shed new light in this direction. On one hand, bulk $MoTe_2$ and $WTe_2$ may host Weyl physics [34, 56] and they also become superconductors at low temperatures and/or under pressure [180-182]. On the other hand, monolayer TMDCs has distinct properties compared to the bulk. Particularly, monolayer $WTe_2$ is found to be a quantum spin Hall insulator with large bandgap [8, 50-52, 54, 55]. Upon moderate electron doping though dielectric field effect, the monolayer also becomes superconductors [183]. Consequently, $WTe_2$ and $MoTe_2$ provide new material platforms to study the interplay between topology and superconductivity, where topological superconductivity and Majorana modes may be possible [22, 23]. More generally, these observations point to the possibilities of using 2D layered materials for identifying and engineering novel topological states of matter and their excitations.

### 6.1. Chiral Anomaly and extremely large magnetoresistance and negative magnetoresistance

In 2014, Ali *et al* [65] reports extremely large, non-saturating magneto-resistance (MR) when a high quality $WTe_2$ crystal is subject to a magnetic field perpendicular to the layer plane (***B***//c axis), as shown in figure 9(a). The MR can be as high as 13 million percent at 0.53 K in a field of 60 T. This unusual behavior highlights the unconventional semimetallic properties of the material. The explanation of the large MR is attributed to the equal populations of electron and hole carriers (electron-hole compensation) in the material [65], which can naturally give rise to the observed quadratic dependence on ***B***. Such explanation in $WTe_2$ is consistent with ARPES measurement on the band structure [184]. Similarly, $T_d$-$MoTe_2$ displays large MR (> $10^6$ %) at low temperatures, too (figure 9(b)) [185-188]. However, ARPES measurement refers $T_d$-$MoTe_2$ as a non-compensated semimetal [188], so that its large MR requires a different explanation. It has also been suggested that spin orbital coupling may play a role in the observed large MR [189]. Understanding this large MR remains to be an open and interesting question.

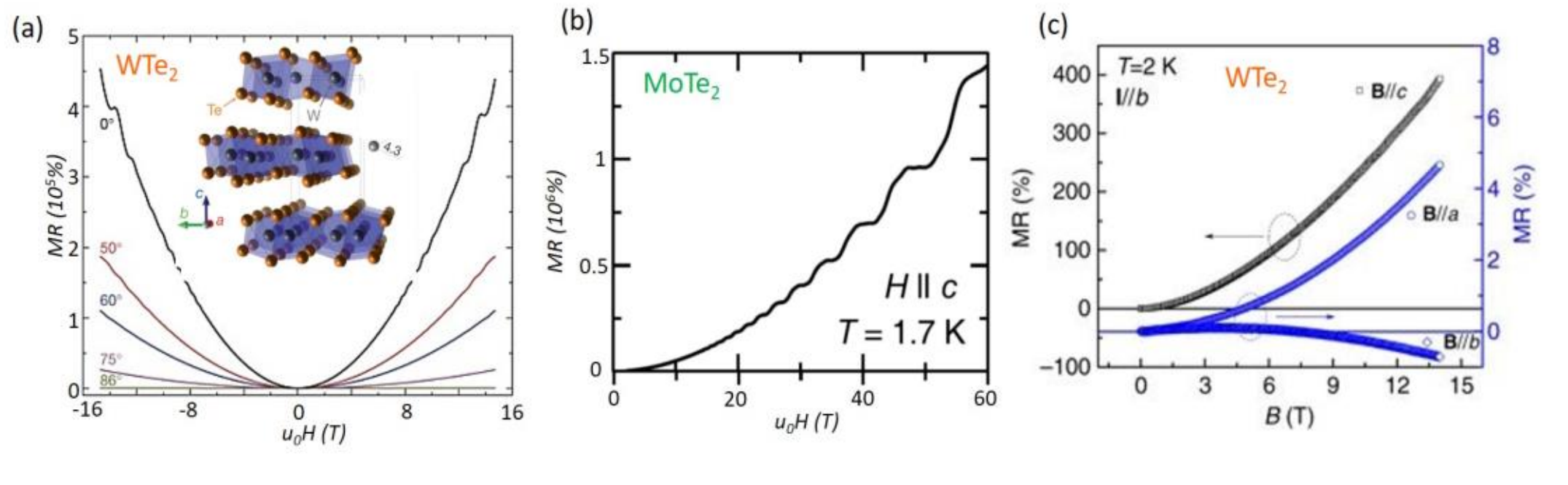

**Figure 9.** Unusual MR in bulk $WTe_2$ and $MoTe_2$. (a) Extremely large magnetoresistance observed in high quality $WTe_2$ crystals. The plot is angular dependence of the MR on the direction of ***B***. $0^o$ refers to ***B***//c while $90^o$ refers to ***B***//a. For this data set, the current is applied along a axis, and $T$ = 4.5 K [65], © 2014 Nature Publishing Group. (b) Similar large MR observed in $MoTe_2$ [187], © 2017 American Physical Society. (c) The MR in $WTe_2$ is strongly anisotropic, and for ***B***//***E***//b, a negative MR is observed, indicating the effect of chiral anomaly [150], © 2017 Nature Publishing Group.

Soon after the observation of the large MR, Soluyanov *et al* in 2005 develops the concept of type II Weyl semimetal [34], of which $T_d$-$WTe_2$ and $T_d$-$MoTe_2$ [56] are the proposed examples. Although experiments have suggested that the Weyl physics is not necessary to the observed large MR under perpendicular fields, the topological physics manifest itself in transport through another remarkable phenomenon called chiral anomaly [190], an interesting quantum behavior that breaks the conservation law of particles with given chirality. The chiral anomaly is first observed in Dirac semimetal $Na_3Bi$ by Xiong *et al* [190]. In their experiment, the observed negative MR for ***B***//***E*** (***E*** is the direction of applied current) is the evidence of the existence of chiral anomaly. Notably, a very recent work has ruled out the trivial explanation of the negative MR based on "current-jetting" effects [191], which further strengthens the conclusion of the chiral anomaly.

Subsequently, the chiral anomaly has been used to test the physics of type II Weyl semimetal predicted in $WTe_2$ [34, 150, 192-196]. Particularly, the tilted Weyl cone in the band structure can lead to an anisotropic negative MR when the direction of ***B***//***E*** is varied, in contrast to the type I un-tilted cone. Indeed, this chiral anomaly evidence has been reported in $WTe_2$ for ***B***//***E***//b axis (figure 9(c)), but not for ***B***//***E***//a axis [150, 196]. This anisotropic effect, together with the observation of an extra quantum oscillation attributed to a Weyl obit, provides an interesting quantum transport study related to the type II Weyl physics [150].

### 6.2. Superconductivity

Interestingly, both $WTe_2$ and $MoTe_2$ are also superconductors. In 2015, two groups report the induced superconductivity in $WTe_2$ under static pressure [181, 182]. The maximum transition temperature $T_c$ is found to be about 7 K at around 13 GPa or 17 GPa (figure 10(a)). A superconducting dome is observed in both studies [181, 182]. The appearance of the superconductivity accompanies with the suppression of the large MR, suggesting the existence of a quantum critical point driven by pressure, as proposed by the authors [181]. The suppression of the large MR can be understood by the violation of the electron-hole compensation due to the structural change of fermi surface. The exact mechanism of the superconductivity is yet confirmed. It has been suggested that a $T_d$ to 1T' phase transition may happen under high pressure [197], which could be related to the observed superconductivity in $WTe_2$. In contrast to $WTe_2$, $T_d$-$MoTe_2$ exhibits intrinsic superconductivity with $T_c$ ~ 0.1 K (figure 10(b)) [180]. High pressure can significantly alter the superconducting properties, and $T_c$ reaches its maximum of about 8.5 K at 11.7 GPa (figure 10(b)). The author also suggests that the 1T' to $T_d$ phase transition may play a role in explanation the superconductivity. No superconductivity has been found under pressure for 2H-$MoTe_2$. These findings point to an interesting situation of superconducting Weyl semimetals, in which topological superconductivity may arise. The topological properties of the superconducting states have yet been studied and their investigation represents an interesting future direction.

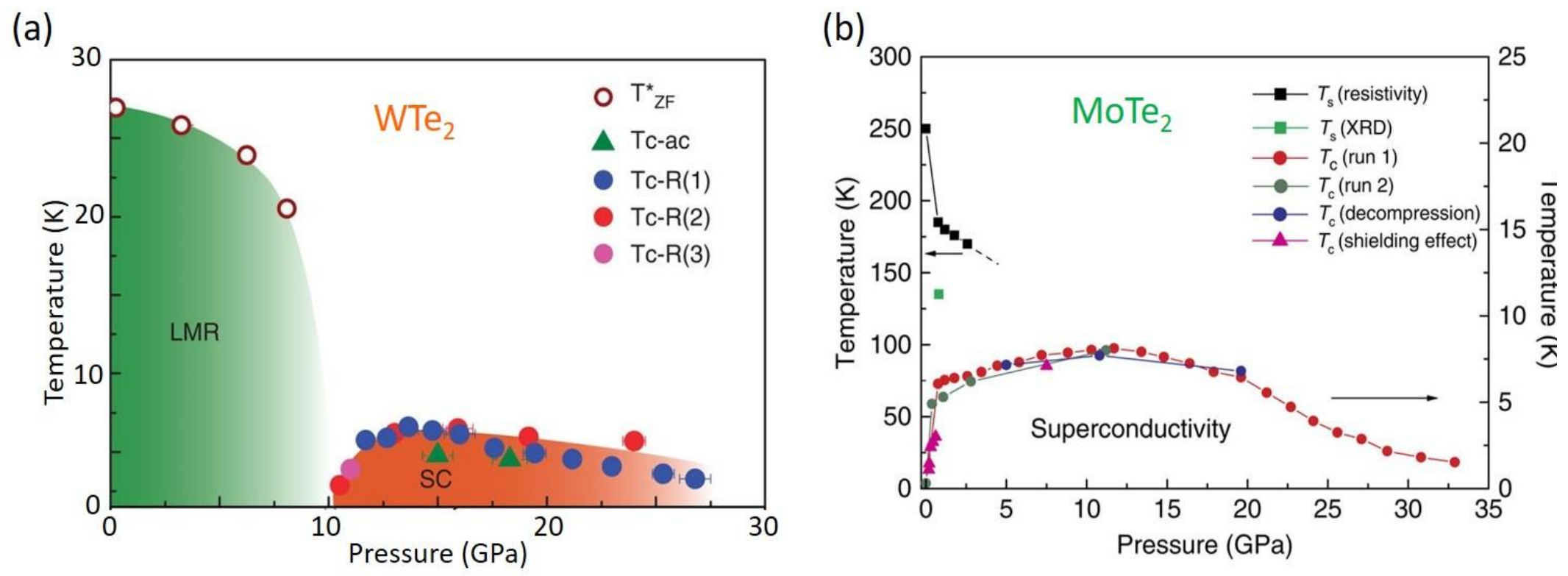

**Figure 10.** Superconductivity in bulk $WTe_2$ and $MoTe_2$. (a and b) Temperature-pressure phase diagrams of $WTe_2$ [181], © 2015 Nature Publishing Group, and $MoTe_2$ [180], © 2016 Nature Publishing Group, respectively.

### 6.3. Quantum spin Hall insulators

The properties of monolayer and few layer van der Waals materials are often distinct from the bulk counterparts. It is also true for $WTe_2$ and $MoTe_2$. Theoretically, it has been predicted that the monolayer 1T' or $T_d$ TMDCs are QSHI if a bandgap develops [8, 50]. Experimentally it has been found that monolayer and bilayer $WTe_2$ indeed acquires a gap (figures 11(a)-(c)) [54], different from the original calculation [8] but consistent with a later one [50]. Few layer $WTe_2$ remains semimetallic (figure 11(a)), but their properties significantly alter from the bulk [198-200]. For example, the MR samples behaves qualitatively different and can be tuned over a large range electrostatically [199]. In few layer $WTe_2$, ferroelectric behaviors appear even at metallic states [201].

The $WTe_2$ monolayer's quantum spin Hall state has recently been intensively investigated using multiple means, including quantum transport [54, 55], ARPES [51], STM [51-53] and microwave impedance microscope (MIM). In quantum transport measurement, the QSH effect produces several remarkable phenomena, which can be used to characterize the effect. First, A QSHI is an insulator in its interior but its edge hosts a conducting channel. Second, the conducting edge channel is described by a helical mode, in which electrons with opposite spin counter-propagates. The helical edge mode, which preserves the time reversal symmetry, results in a highly specific quantum conductance of one $e^2/h$ per edge. Third, the edge conductance should be significantly reduced under breaking time reversal symmetry due to the loss of protection.

Experimentally, Fei *et al* [54] showed that the edge state conduction appears in monolayer but is absent in bilayer. The observed conductance of monolayer edge is less than the expected quantum value and is associated with a zero-bias anomaly, which prevents the identification of the helical nature. By designing a specific device geometry combing bottom and top gates, Wu *et al* [55] successfully revealed the intrinsic edge conductance of monolayer $WTe_2$ to be ~ $e^2/h$ per edge (figure 11(d)), confirmed by length dependence study. The device's behavior under magnetic fields is also consistent with the prediction from QSH effect. Particularly, the authors observed evidence of the

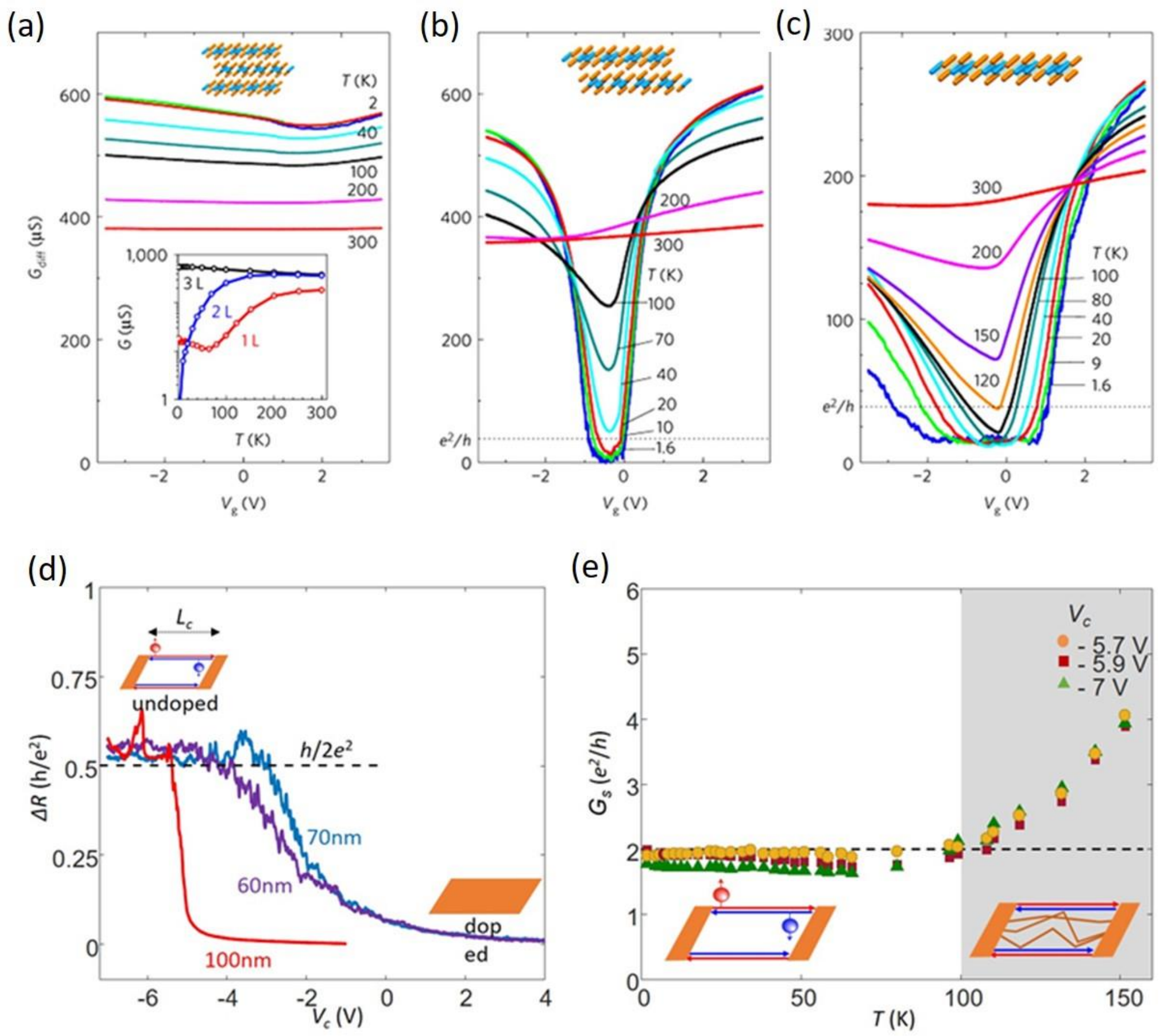

**Figure 11.** Quantum spin Hall effect in monolayer $WTe_2$. (a-c) Gate dependent conductance of measured tri-, bi- and mono-layer flakes of $WTe_2$, taken under different temperatures. The trilayer behaves as a semimetal. Monolayer and bilayer, however, behaves like an insulator near zero doping. Highly doped monolayer or bilayer recovers to be metallic [30], © 2017 Nature Publishing Group. (d) Measurement of the resistance of short monolayer channels defined by gates. The plateau appears due to the presence of the edge state channel and its value quantifies the edge channel resistance. For three channels with different length (indicated by the numbers in nanometers) and width, the plateaus all appears to be of the same value, $h/2e^2$ for two edges. (e) The resistance of the 100 nm channel stays near the quantum value up to 100 K, indicating the high temperature QSH effect. (Figures (d) and (e) are adapted from [55], with permission from AAAS).

Dirac point in the helical edge band. Combined with theory and other types of measurements, the QSH state in monolayers $WTe_2$ is believed to be established. Due to the large band gap, the QSH state dominates the transport up to ~ 100 K (figure 11(e)), significantly larger than the other two QSH systems based on semiconductor heterostructures.

### 6.4. Field effect induced superconductivity

Superconductivity also appears in the monolayer [183]. When the same QSH device reported in Ref [55] is cooled down to below 1 Kelvin, a superconducting transition is found when the $WTe_2$ monolayer is electron doped by the gates (figure 12(a)). The same result has also been observed by Cobden and Folk's collaboration. [202] The critical doping for the onset of superconductivity is reported to be about $5 \times 10^{12}$ $cm^{-2}$, a very low density for superconductors. $T_c$ increases with increasing doping in the accessible region of the reported devices, with a maximum about 1 Kelvin. Moreover, the upper critical field significantly exceeds the Pauli paramagnetic limits when in-plane magnetic field is applied. The reason is yet understood. Note that the superconductivity in bulk $WTe_2$ is found only by applying pressure and few layer $WTe_2$ grown by MBE also displays superconductivity beyond Pauli limit [200].

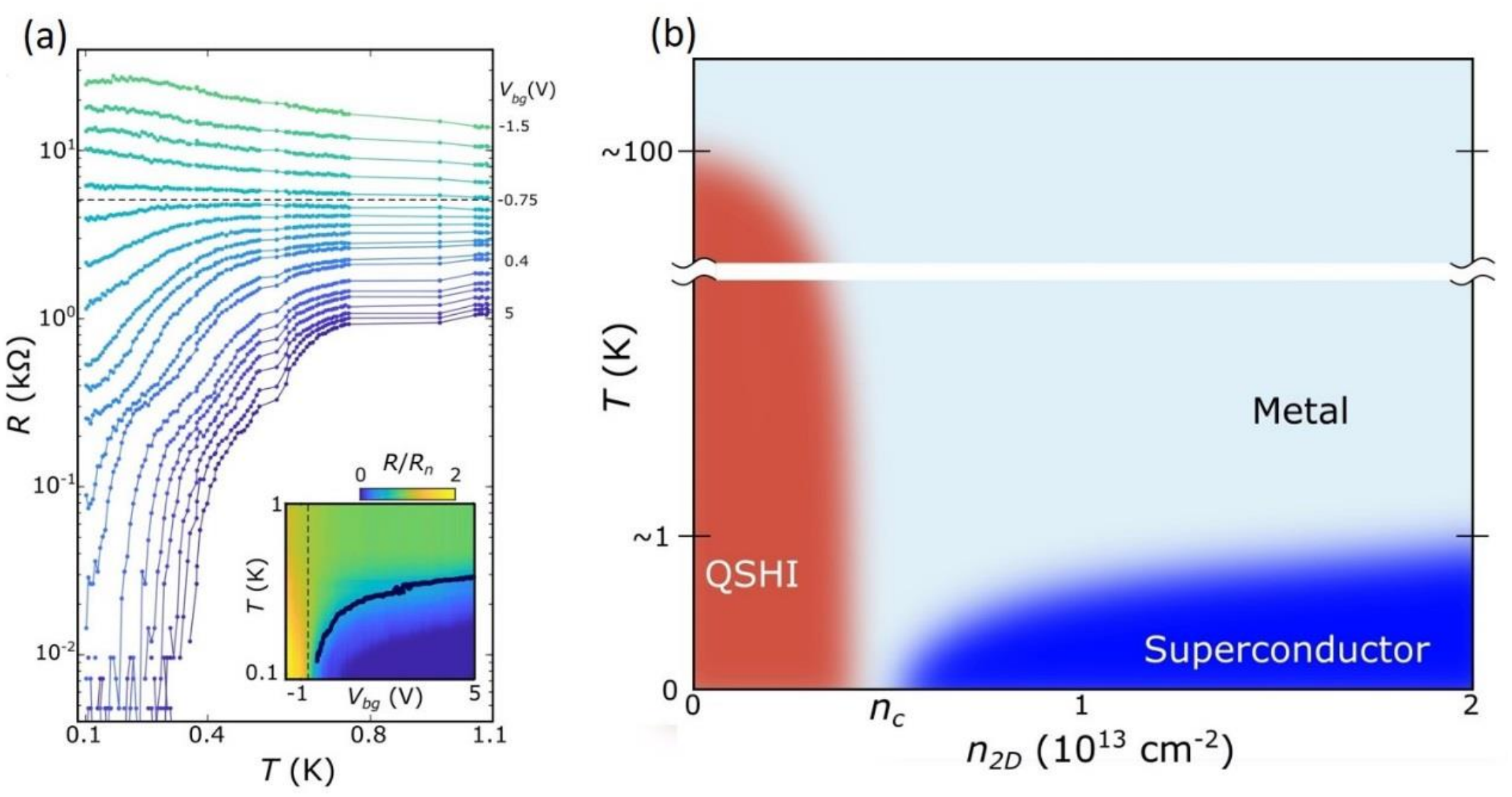

**Figure 12.** Superconductivity in monolayer $WTe_2$. (a) Gate dependent resistance as a function of temperature for the monolayer flake. Inset shows the same data but in color plot. Superconductivity appears after a critical electron doping. (b) Electronic phase diagram of monolayer $WTe_2$ [183].

A preliminary electronic phase diagram for monolayer $WTe_2$ is summarized in figure 12(b), in which a QSH insulating phase resides nearby the superconducting phase. Interesting future directions include the investigation of (1) the origin of the insulating and superconducting gap; (2) the topological properties of the superconducting state; and (3) the possibilities of constructing non-abelian excitations in the monolayer. Moreover, the electric field induced superconductivity in the monolayer allows for engineering superconducting nano-device in a single atomic plane. It will also be exciting to extend the study to monolayer $MoTe_2$ and other van der Waals monolayers and heterostructures.

## 7. Optical Response

Optical approaches are indispensable routes to probe the fundamental topological properties and to control of electron's quantum degree of freedoms. Although there are many theoretical proposals toward this direction on topological semimetals, the experimental progress lacks behind because the relevant optical wavelength range to probe the vicinity of the Dirac/Weyl points of the topological semimetals usually lies in the THz to mid-IR wavelength range which is technically more challenging comparing to relatively matured visible and near-IR wavelength range. Regarding mechanically exfoliated 2D layered materials, it is more difficult due to the limitation on the size of the sample that can be obtained, which is usually smaller than the diffraction limit of the desired wavelength. In this part, we review the current optical experiments progress in verification of topological semimetals and interesting optical phenomenon that can manifest the topological features of topological semimetals. As most experimental progress so far is based on bulk topological semimetals instead of 2D layered ones, the experimental results discussed in this session are mostly based on bulk materials, especially model systems such as $Cd_3As_2$ and TaAs, but the optical effects discussed below are quite universally applicable to 2D layered topological semimetals. The related practical optical device application prospects are also discussed.

### 7.1. Linear ac optical interband response

The simplest Hamiltonian that describes a semimetal with energy directly proportional to the crystal momentum via an isotropic Fermi velocity would lead to an ac optical interband response linear with photon energy($\Omega$), and passes through the origin in the limit $\Omega \rightarrow 0$ [203]. This linear behavior is widely considered to be an important optical signature of 3D Dirac Semimetal. While in the Weyl phase, due to the break of the degeneracy of Dirac cones, instead of a linear response, there are two quasilinear regions with different slopes in the interband optical response as shown in figure 13(a) [204]. A singularity is proposed to occur at $\Omega_c$. Transitions occurring in the region between the two Weyl nodes are blocked when $\Omega$ exceeds $\Omega_c$, giving rise to the reduced slope. The evolution of ac optical interband response from Dirac Semimetal to Weyl semimetal and gapped semimetal phases are discussed theoretically in detail by C. J. Tabert and J. P. Carbotte [204] and S. P. Mukherjee and J. P. Carbotte [205]. In real material, the Dirac/Weyl cones will have certain tilting and doping, this linear relationship is modified as the cone is titled or doped, which makes the verification of Dirac/Weyl semimetal through an optical conductivity measurement more intriguing. The effect of tilting and doping on the optical response of a Type-I and Type-II Weyl semimetals are also systematically treated by J. P. Carbotte [206, 207]. Experimentally, the linear interband ac optical response has already been verified through Fourier transform infrared (FTIR) spectroscopy in some works, such as on layered Dirac Semimetal candidate $ZrTe_5$ [208], Weyl semimetal candidate pyrochlore $Eu_2Ir_2O_7$ [209], Type-I Weyl semimetal TaAs [210, 211] and Type-II Weyl semimetal $WTe_2$ [212]. However, some other FTIR measurement results show discrepancies from linear ac optical response, which are attributed to doping and intraband response [209, 213]. After all, it is questionable sometime whether the investigated material itself is indeed Dirac/Weyl semimetals [214]. Theoretically, although the bulk state usually dominates the response, Li-kun Shi and Justin C. W. Song [215] noted the Fermi arc surface states can possess a strong and anisotropic light-matter interaction, characterized by a large Fermi arc optical conductivity when light is polarized transverse to the Fermi arc as shown in figure 13(b). This large and anisotropic Fermi arc conductivity is proposed to be a novel means to optically interrogate the topological surfaces states of Dirac semimetals, but remain to be elusive experimentally.

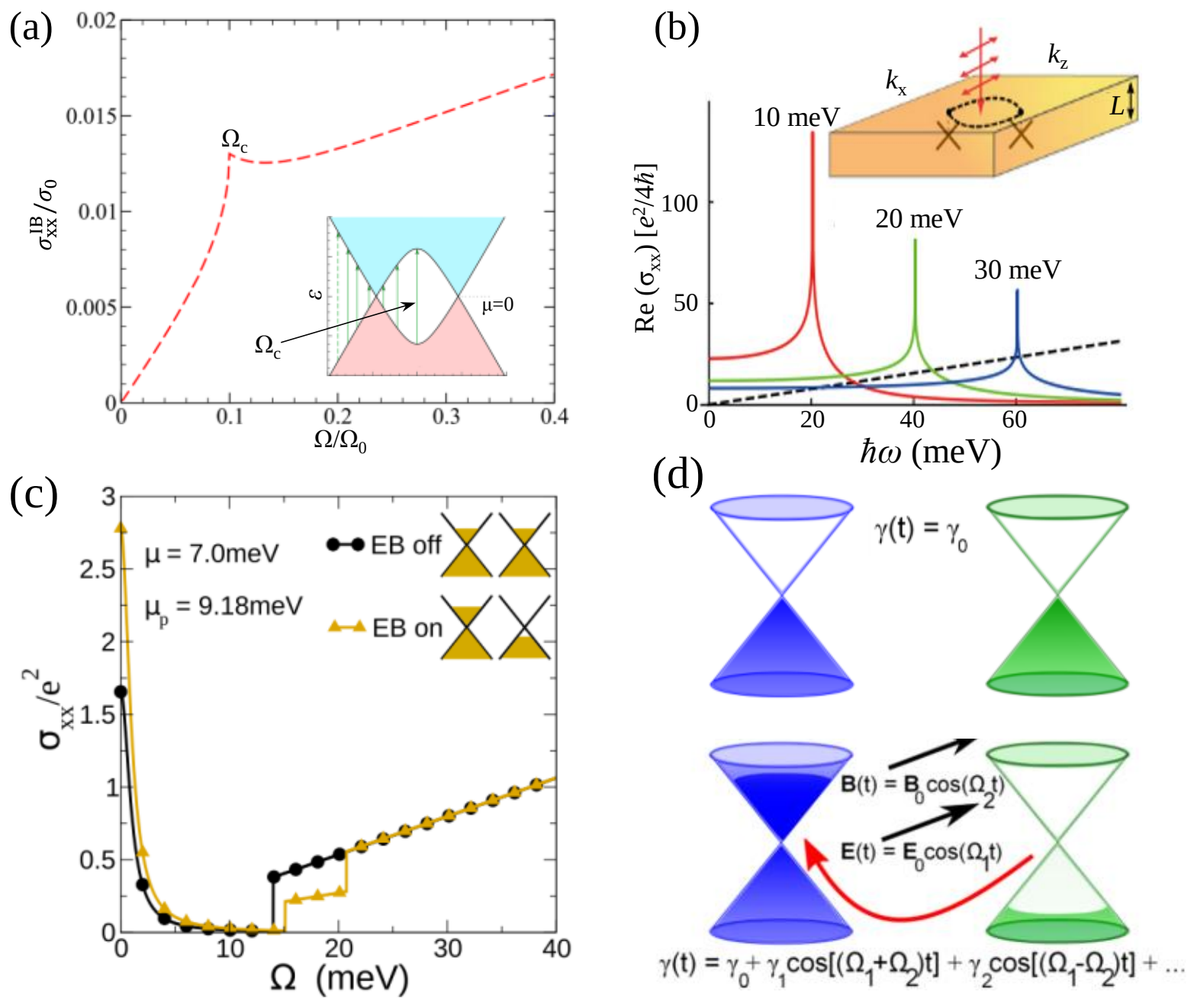

**Figure 13.** Calculated interband optical conductivity ($\sigma_{xx}$) and chiral anomaly. (a) $\sigma_{xx}$ in the WSM phase before the massive band contributes. Two quasilinear regions are observed with a singularity occurs at $\Omega_c$. Inset shows the band structure near the Weyl nodes (adapted from [204], © 2016 American Physical Society). (b) $\sigma_{xx}$ with a distinctive peaklike frequency dependence arising from the DSM Fermi arc surfaces states (adapted from [215], © 2017 American Physical Society). (c) $\sigma_{xx}$ with steplike signatures for a clean Weyl semimetal after the application of ***E B*** field. (yellow) [217], © 2014 American Physical Society. (d) Illustration of gyrotropy induced by an ***E B*** field in WSMs. Colored (white) regions denote filled (empty) states and the two colors indicate Weyl nodes of opposite chiralities. The curved red arrow shows charge transfer across the nodes pump by an ***E B*** field, resulting in a charge imbalance between the nodes and a consequent net anomalous contribution to γ (adapted from [218], © 2015 American Physical Society).

### 7.2. Chiral Anomaly related optical response

If further asked whether there is intrinsic topological signature of Dirac/Weyl semimetals that can be manifested in optical spectrum, the answer may be related to chiral anomaly, which is one of the remarkable characters of these systems Mechanism of chiral anomaly has been discussed in the section 6.1. Many early attempts to detect Dirac/Weyl semimetals have focused on transport experiments regarding chiral anomaly. Optically, so far few experimental evidence has been established, such as a large Kerr rotation in $Cd_3As_2$ [216], but many theoretical schemes have been proposed regarding this aspect. Ashby and Carbotte have proposed an optical absorption experiment, in which the interband optical conductivity shows step-like features at finite frequencies as shown in figure 13(c) [217]. However, low temperature and high quality samples are suggested for measurable experimental features, as high temperature and scattering processes caused by impurities can smooth these features out. Additionally, the photon energy is below 40 meV in their calculation, so the absorption experiment has to be carried out in THz regime. Alternatively, Hosur and Qi *et al* proposed that the chiral anomaly can be probed by measuring the optical activity via circular dichroism. Particularly, in the absence of $\boldsymbol{E}\cdot\boldsymbol{B}$, the Fermi levels at the two Weyl nodes with opposite chiralities are equal. An applied $\boldsymbol{E}\cdot\boldsymbol{B}$ field pumps charge across the nodes, resulting in a charge imbalance between them and a consequent net anomalous contribution to gyrotropic coefficient ($\gamma$). Since WSMs break time reversal or inversion symmetry, they are in general optically active in the absence of external fields as well. The anomalous contribution, proportional to $\boldsymbol{E}\cdot\boldsymbol{B}$, can be isolated by applying electromagnetic fields at distinct finite frequencies. Apart from the nonzero $\gamma$ (figure 13(d)), a Hall-like conductivity can emerge as well, both of which may be detectable by routine circular dichroism experiments. This method also serves as a diagnostic tool to discriminate between Weyl and Dirac semimetals; the latter will give a null result [218]. However, Weyl semimetals can exhibit optical activity due both to its topological electronic structure and to crystallography. The effect from crystallography has to be carefully ruled out by constructing experiments where this effect is minimized, so that the unique topological signatures can be identified [219]. Regardless, the novel Weyl semimetal TaAs, as well as its related siblings (TaP, NbAs, NbP), should exhibit novel optical activity of a polar vector nature that in principle can be identified by appropriate resonant x-ray diffraction measurements, as proposed by Norman *et al* [219].

### 7.3. Broadband Photocurrent Response

When it comes to the interplay of the optical and electrical response, the measurement of photocurrent response is an effective experimental approach. When light excitation is above the bandgap, electron-hole pairs will be injected in the material and separated by certain mechanisms, which lead to measurable photocurrent in external circuit. In topological Dirac/Weyl semimetals, the linear energy dispersion and zero bandgap are symmetry protected around Dirac/Weyl nodes, allowing photons of arbitrary energy to excite the electrons from valence to conduction band if the Fermi level is at the Dirac/Weyl nodes with no Pauli blocking effect due to doping. Thus the photo-response of Dirac/Weyl semimetals are usually broadband and suitable for photo-detection over full broad spectrum range (figure 14(a)). The broadband response from visible to mid-infrared has been verified by Wang *et al* [220] and Lai *et al* [221, 222] on 3D Dirac Semimetal $Cd_3As_2$ and layered type-II Weyl semimetal $MoTe_2$ and $TaIrTe_4$ respectively. These photodetectors can work in an unbiased self-powered mode as a result of zero bandgap feature of semimetals and the unbiased operation can help avoid the dark current issue that always exists in a biased photodetector. With a cross-polarized pump-probe measurement, the response time of $Cd_3As_2$ device is approximately 6.87 ps (figure 14(b)), which is comparable to graphene [220]. In contrast, the response time of photodetectors based on layered type-II WSMs $MoTe_2$ and $TaIrTe_4$ are approximately 31.7 μs and 27.0 μs under 10.6 μm excitation [221-223]. In addition, these layered WSMs show strong polarization dependence. The anisotropy is found to be wavelength dependent and increase as the excitation gets closer to the Weyl nodes.

### 7.4. Transient optical response and practical ultrafast application prospects

Dirac/Weyl semimetals feature gapless band structure and rapid transient relaxation time of photoexcited carriers, which enables convenient optoelectronic devices with ultrafast response and broad spectrum range. In terms of immediate practical optoelectronic applications, saturated absorber and optical switch are two prospects beyond photodetection based on DSM/WSM. Among various DSM/WSM, the transient dynamics of $Cd_3As_2$ has been very well characterized with broadband probe from near-infrared to THz [224-226], indicating transient time on the order of picosecond timescale (figure 14(c)). Transient spectroscopy work on other DSM/WSM indicates similar transient time scale [227]. This transient time is similar to graphene and quite typical for semimetals. Benefiting from this ultrafast transient time, ultrafast photodetector, optical switch and saturate absorber (figures 14(d)-(g)) based on $Cd_3As_2$ has realized experimentally, and among them, the photo-detection and optical switch is demonstrated to work in mid-infrared range [220, 228, 229].

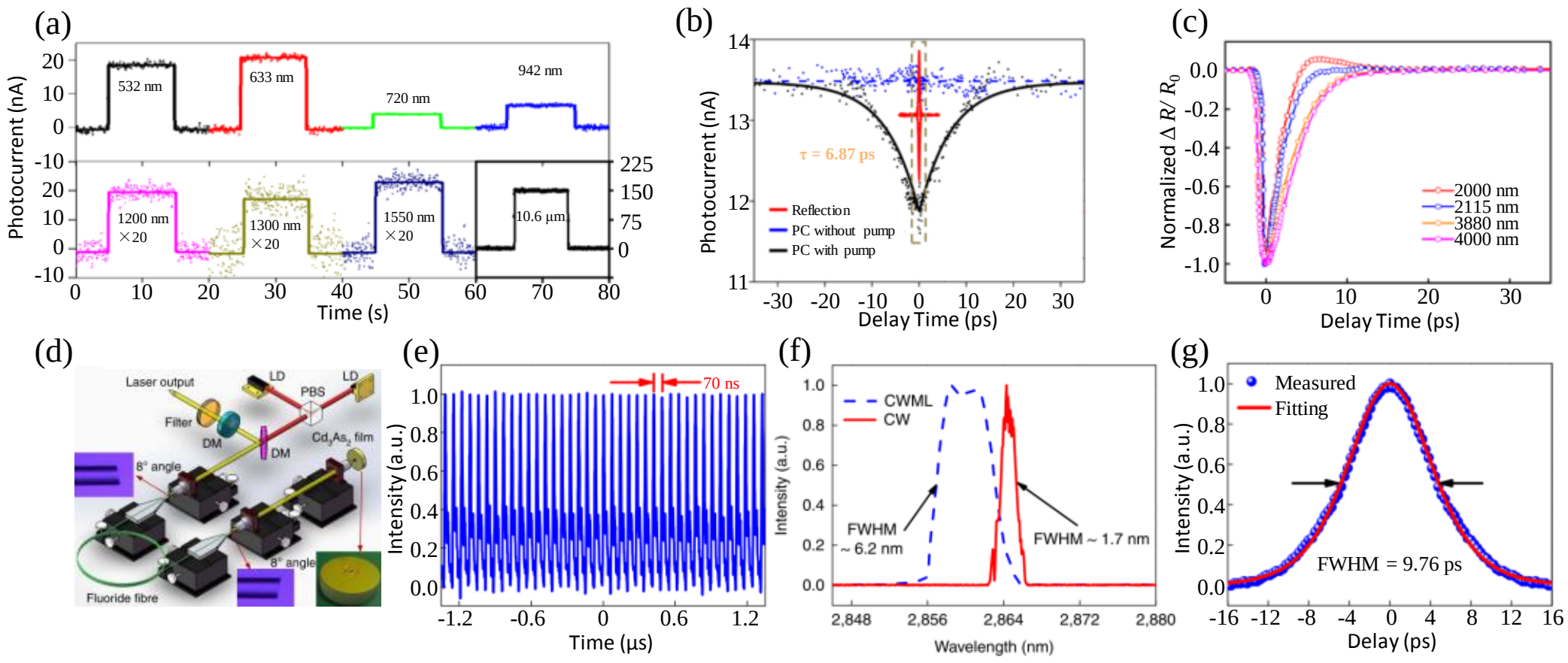

**Figure 14.** (a) Photoresponse of a $Cd_3As_2$ photodetector over broad wavelength range from visible to mid-IR. (b) Ultrafast time resolved measurement of photo response time of $Cd_3As_2$ photodetector. Probe induced photocurrent with pump on (black) and off (blue) as a function of pump−probe pulse delay. The results shows a response time of 6.87 ps. Figures (a) and (b) are reproduced with permission from [220], © 2017 American Chemical Society. (c) Probe wavelength dependence of transient dynamics on bulk $Cd_3As_2$ with pump wavelength of 800 nm at 77 K [225], © 2017 American Physical Society. (d) Schematic of ultrafast mid-infrared fiber laser based on $Cd_3As_2$ saturable absorber. (e) CW mode-locked pulses at a pump power of 286.9 mW, indicating stable operation of the mid-infrared fibre laser. (f) Output optical spectrum of the mid-infrared fibre laser. (g) Autocorrelation trace of the mid-infrared fibre laser with extracted pulse width of 6.3 ps. Figures (d)-(g) are reproduced with permission from [228], © 2018 Nature Publishing Group.

### 7.5. Circular Photogalvanic Effect and Optical excitation of Chiral carriers

Weyl nodes are monopoles of Berry curvature and low energy excitations in the vicinity of Weyl nodes behave as chiral Weyl fermions. The chirality of Weyl points in WSM brings abundant physical phenomena that correlates to circular dichroism according to the optical selection rule derived from the conservation of angular momentum: the absorption of left/right circular polarized (LCP/RCP) light couples to the Weyl points with opposite chirality as shown in figure 15(a). Circular dichroism is an important tool to detect and control chirality in Weyl semimetals. These aspects have attracted broad theoretical and experimental attentions, as this would lead to experimentally measurable effect that is related to the chiral nature of the carriers in Weyl semimetals. In single Weyl cone, circular polarized light is supposed to inject carriers on one side of the Weyl cone, leading to directional current. In Dirac semimetals (DSM), two Weyl points with opposite chirality are degenerate and form one Dirac point. Therefore, the circular selection rules vanish in Dirac semimetal as transitions are allowed on both side of the Dirac cones with circular excitation [230].

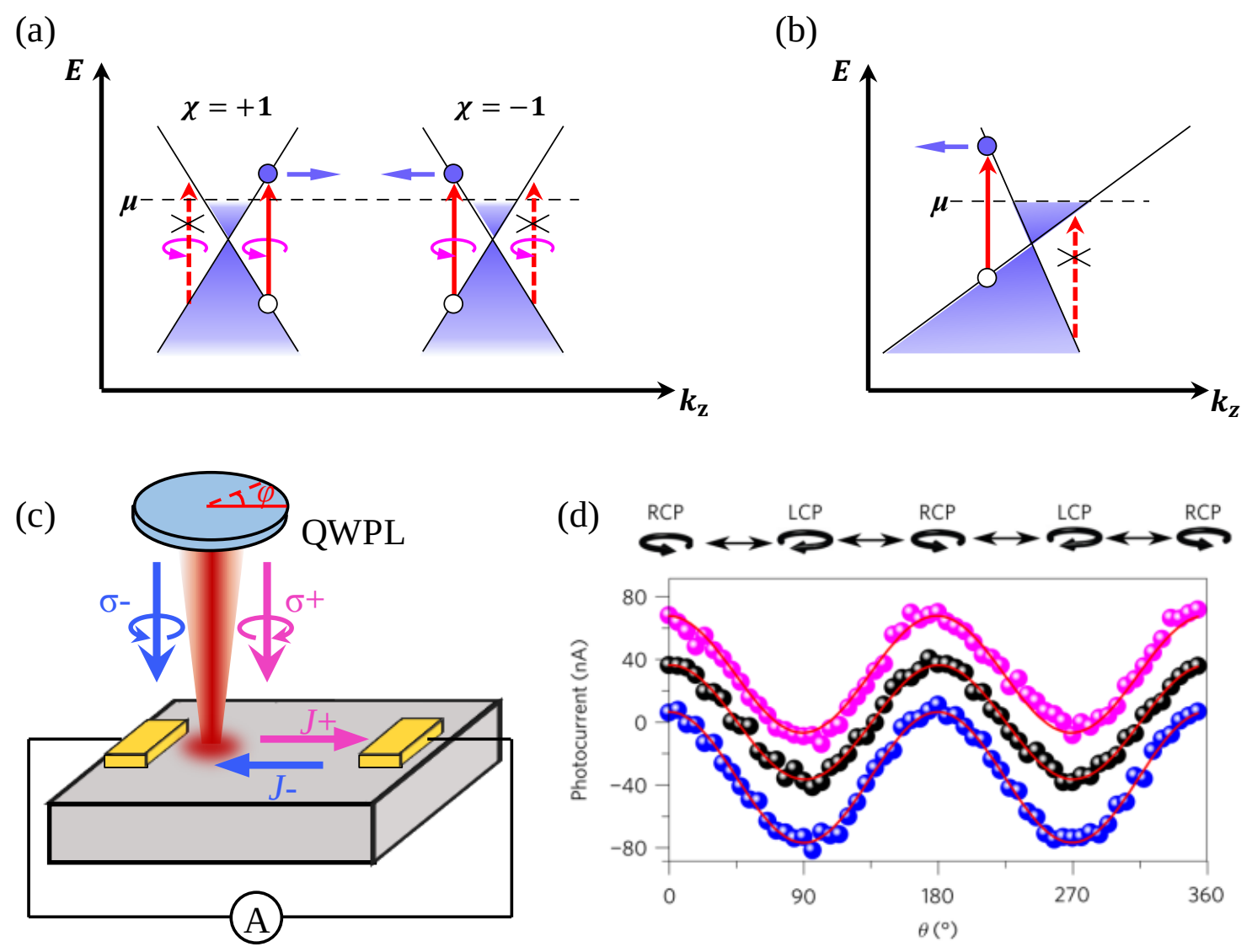

**Figure 15.** (a) Chirality selection rule: LCP (RCP) light along $+\hat{\mathbf{z}}$ excites the $-k_z$ ($+k_z$) side of the $\chi$ =+1 Weyl node but the $+k_z$ ($-k_z$) side of the $\chi$ =−1 Weyl node. (b) In the presence of a finite tilt and a finite chemical potential away from the Weyl node, the Pauli blockade becomes asymmetric about the nodal point, which result in non-cancelled photocurrent [230, 231]. (c) Schematic illustration of the mid-IR photocurrent microscope setup. Polarization is tunable between LCP and RCP by rotating the quarter waveplate (QWPL) and the direction of the photocurrent is tunable as a result. (d) Polarization-dependent photocurrents at $T$=10 K measured along specific orientation of TaAs with 10.6 μm excitation [231], © 2018 Nature Publishing Group.

Similarly, as the Weyl cones with opposite chirality always come in pairs in any WSMs, the directional current generated within this pair can cancel each other, providing no experimentally measurable net photocurrent. However, Chan *et al* proposed that this problem can be addressed in WSMs with tilted Weyl cones as a result of the decreased symmetry [230] as shown in figure 15(b). Combined with Pauli blockade and circular selection rules, the optical transition in one Weyl cone is blocked, which results in directional photocurrent generation under the excitation of circular polarized light. As LCP and RCP light can inject photocurrent with different direction (figure 15(c)), this would lead to nonvanishing Circular photogalvanic effect (CPGE), which is observable experimentally through circular polarization dependent photocurrent measurement by rotating a quarter waveplate. CPGE has been experimentally observed by Ma *et al* [231] on Type-I Weyl semimetal TaAs along specified crystal direction with 10.6-μm excitation, where the direction of the photocurrent changes depending on the polarization of excitation light as shown in figure 15(d).

In layered type-II WSMs, such as $MoTe_2$ and $TaIrTe_4$, although the Weyl cones are strongly tilted, the in-plane $C_{2v}$ symmetry leads to cancellation of CPGE effect of a pair of mirror symmetry related

Weyl cones. However, the CPGE effect is also experimentally observed in these materials [232, 233]. Certain mechanisms has to come into play to break the mirror symmetry to provide a net CPGE, which are still under debates. Ji et. al. observed CPGE with 750 nm excitation in layered type-II WSM $MoTe_2$. The $C_{2v}$ symmetry is claimed to be broken as a result of spatially inhomogeneous optical excitation due to a focused Gaussian beam in this work. Nevertheless, the optical excitation in this work is far from the Weyl cones, and the physics behind it may not related to the chirality of Weyl cones [232]. In parallel, Ma *et al* [233] also observed CPGE in $TaIrTe_4$ with 4.0 μm and 10.6 μm excitation, where this symmetry breaking is attributed to a background of DC electrical field generated by photo-thermoelectricity or built-in electric field at the interface with electrodes.

When it thins down to monolayer, as QSHI discussed in section 5.3, strong and electrically tunable Berry curvature concentrate near the band edge and show a bipolar configuration about the mirror plane. As a result, interband transitions near the band edge also select opposite circular polarization light, while CPGE occurs when a nonzero Berry curvature dipole exists. Recently, with monolayer QSHI $WTe_2$, this effect has been observed experimentally by Xu *et al* [234].

### 7.6. Topological Enhanced Nonlinear Response

When strong light irradiation is applied to materials, nonlinear polarizations happen under strong ac electric field and realize abundant nonlinear optical effects. In topological semimetals, by scaling laws, the Dirac fermion in two dimensions shows the giant nonlinear responses to electromagnetic fields in the terahertz region as shown in figure 16(a) [235]. Besides, various nonlinear optical effects, such as the shift current, photovoltaic Hall response, are governed by a vector field defined by Berry connection, so that they “can be described in a unified fashion by topological quantities involving the Berry connection and Berry curvature” theoretically [236] as shown in figures 16(b) and (c). As the Berry curvature diverges at the Weyl nodes, the nonlinear response is usually enhanced at the singularity points as shown in figures 16(d) and (f) [237, 238].

In presence of magnetic field, the linear and second-order nonlinear optical responses of Weyl semimetals are theoretically treated by a semiclassical Boltzmann equation approach by Takahiro Morimoto *et al* [239]. The theory predicts strong second harmonic generation (SHG) proportional to $\boldsymbol{B}$ that is enhanced as the Fermi energy approaches the Weyl point, as a consequence of the divergence of the Berry curvature and orbital magnetic moment near the Weyl point. In this regard, the SHG of Weyl semimetals under $\boldsymbol{B}$ is tied to the monopole physics in the momentum space described by the Berry curvature. Especially, when the magnetic field is parallel to the optical electric field, chiral anomaly enhances the SHG signal compared to the case where $\boldsymbol{B}\perp\boldsymbol{E}$. In practice, these divergences are cut off by the energy broadening due to the nonzero relaxation time and by excessive electric field $\boldsymbol{E}$ as the semiclassical treatment is invalid then. Experimentally, the enhancement of SHG can be detected as a large Kerr rotation signal, and the magnitude of the nonlinear magneto-optical susceptibility of TaAs at 0.1 eV photon with magnetic field is estimated to be 2-3 order larger than GaAs at visible light range [239].

The experimental observation of such strong SHG is first realized by Wu et. al [240] on type-I WSM TaAs. Strikingly, an extreme anisotropy and large absolute magnitude of second-harmonic optical susceptibility were detected from (112) surface without an external magnetic field, which can exceed the values in the benchmark materials GaAs and ZnTe by approximately one order of magnitude as shown in figure 16(e). The measurement, however, is performed with fundamental wavelength of 800 nm, which is far above the transition around the singularity point (Weyl node). As a result, connection between large nonlinear coefficient and the Weyl physics remains to be investigated further with low photon energy measurement at mid/far-infrared or even THz wavelength range. Experiments on Type-II category regarding SHG is still missing.

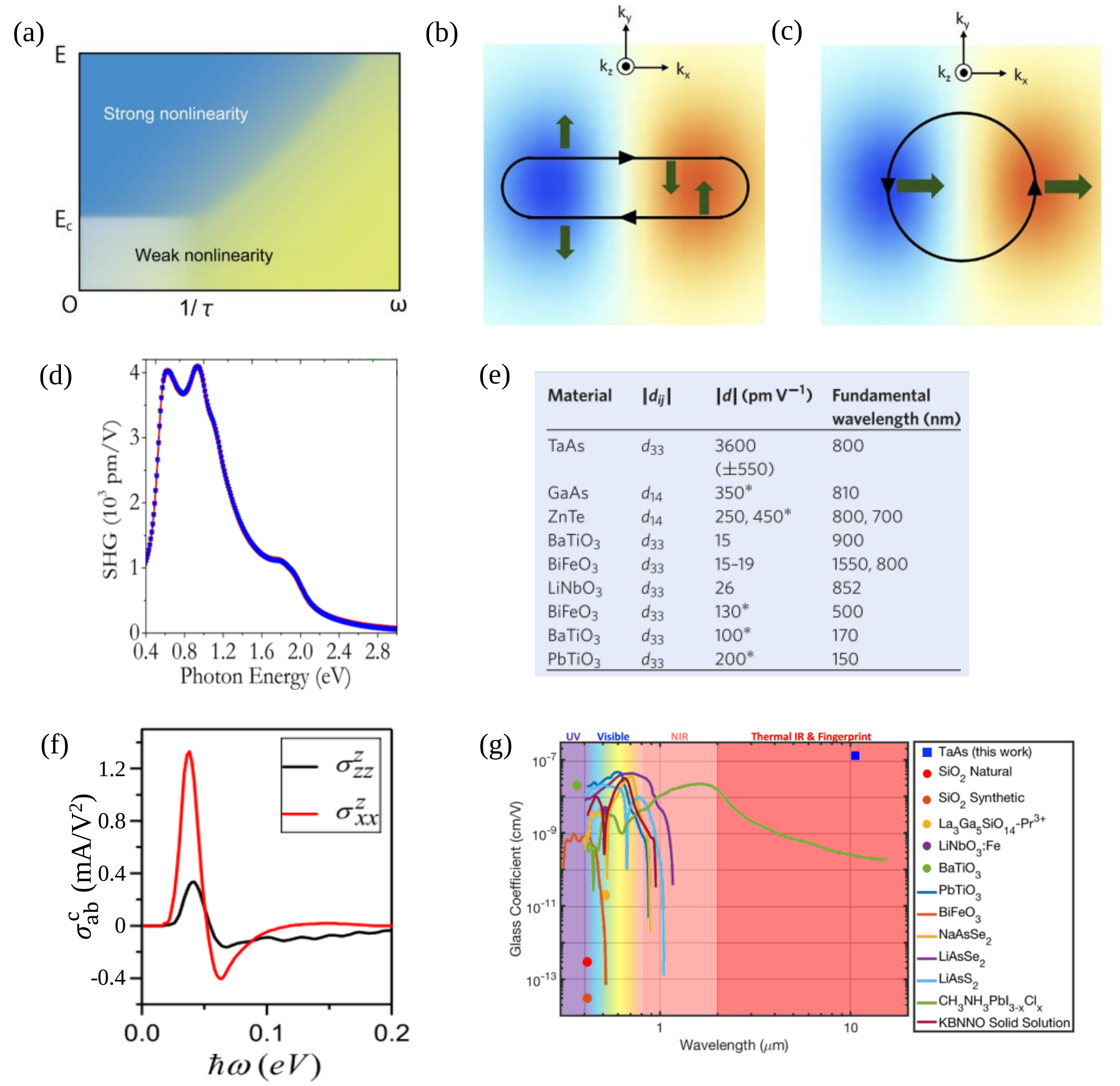

| Material | $\lvert d_{ij}\rvert$ | $\lvert d\rvert$ (pm V$^{-1}$) | Fundamental wavelength (nm) |
|---|---|---|---|
| TaAs | $d_{33}$ | 3600 (±550) | 800 |
| GaAs | $d_{14}$ | 350* | 810 |
| ZnTe | $d_{14}$ | 250, 450* | 800, 700 |
| $BaTiO_3$ | $d_{33}$ | 15 | 900 |
| $BiFeO_3$ | $d_{33}$ | 15-19 | 1550, 800 |
| $LiNbO_3$ | $d_{33}$ | 26 | 852 |
| $BiFeO_3$ | $d_{33}$ | 130* | 500 |
| $BaTiO_3$ | $d_{33}$ | 100* | 170 |
| $PbTiO_3$ | $d_{33}$ | 200* | 150 |

**Figure 16.** (a) Two regimes of nonlinear responses in the $\omega$-$E$ plane as a result of Dirac fermions [235], © 2016 American Physical Society. (b), (c) Semi-classical picture of SHG and CPGE induced by a Berry curvature dipole. The distribution of Berry curvature in momentum space is indicated by the color scale [239], © 2016 American Physical Society. (d) Calculated SHG coefficients $\chi^{(2)}_{zzz}$ (2ω; ω, ω) contributed from shift current of TaAs [237], © 2018 American Physical Society. (e) SHG coefficients of different materials at room temperature [240], © 2018 Nature Publishing Group. (f) Photon energy dependence of shift current conductivity under linearly polarized light in Weyl semimetals from first principles [238], © 2018 American Physical Society. (g) Experimentally measured (circles) and calculated (lines) Glass coefficient for various ferroelectric materials [223].

Another Berry curvature singularity enhanced nonlinear response in WSMs is the shift current response. This has been observed experimentally in both Type-I TaAs by Osterhoudt *et al* [223] and Type II $TaIrTe_4$ by Ma *et al* [233]. The measurement on TaAs by Osterhoudt *et al* [223] is performed with single wavelength excitation at 10.6-µm, with a glass coefficient (namely responsivity divided by the absorption coefficient) reported nearly one order of magnitude larger than previous results, as shown in figure 16(g). In Type-II WSM $TaIrTe_4$, Ma et. al. has observed unusual large photocurrent response with 4-µm excitation, which is attributed to third order nonlinear response instead of second order, as the related second order nonlinear coefficient should be zero due to mirror symmetry of the material [233]. Multiple wavelengths are measured in this work, and the exceptional responsivity observed at 4 µm is believed to related to the doping level of the sample as predicted by theoretical work of Xu *et al* [241]. Indeed, Xu *et al* also point out theoretically that the shift current response has strikingly different frequency dependent behaviors between type-I and type-II WSMs in terms of their frequency dependence. The optical conductivity due to shift current response $\sigma_{\mathrm{shift}}$ of type-I WSMs is proportional to $\omega$ under zero doping and zero temperature, exhibits a vanishing property. In stark contrast, in type-II WSMs, second-order optical conductivity $\sigma_{\mathrm{shift}}$ is inversely proportional to the frequency $\omega$ of incident light under zero doping and zero temperature. Though, the vanishing and divergent behaviors will be truncated and a peak at tunable frequencies with tunable amplitude will be formed with doping and temperature changed [241] Concretely, shift current calculation in the realistic material band structures of TaAs was also carried out by Zhang *et al* recently and showed a similar result [238]. To conclude, the diverging property originates from monopoles of the Berry connection around Weyl nodes and results in strongly enhanced photocurrent response under long wavelength excitation. Inversely, we note the shift current response also provides a means to probe the Berry curvature and distinguish Fermi surface topologies from this aspect.

## 8. Outlook

Existing experiments have focused on the discovery of new topological materials and the demonstration of the associated physics. In material aspects, stable Weyl semimetals with minimum number of nodes separated from trivial bands, ideally close to the Fermi level with a large momentum separation between nodes, are highly desired. In layered category, $TaIrTe_4$, as an inversion symmetry breaking Weyl semimetal [39, 242, 243], may be a good candidate. In addition, exotic types of WSM, such as a magnetic WSM that breaks time reversal symmetry but preserves inversion symmetry, is still lacking and its layered candidates are highly desired. In long term, synthesis of large-area high-quality materials is always prerequisite for not only the demonstration of interesting topological physics but also mass scale device applications. For layered materials, it is equally important to achieve the epitaxial growth of multiple layers with atomically clean interface between different layers to form various kinds of heterostructures.

Innovating measurement and control schemes are also essential to move forward. In fact, many exotic quantum states cannot be well understood without the development of new probing tools. These new tools may require the integration of different kinds of techniques at extreme conditions. Achieving quantum control of the new physics is another direction, which may be possible by integrating electrical and optical means, leading to practical applications. Particularly, it is desirable to develop techniques at different frequencies, especially in GHz and THz range, tailored for the interesting topological quantum physics. Straightforwardly, topological semimetals can be employed in high-speed electronics, optoelectronics and spintronics, exploiting the high mobility and large MR [65, 185-188, 244-246]. Another possibility for topological insulators and Weyl semimetals is to utilize their robust surface states for surface-related chemical process such as the catalysis [247].

The unique, unprecedented advantages of 2D layered materials lie in its easy access to interface engineering, by forming Van der Waals heterostructures [248-250]. Such structures can integrate superior functionality of established 2D layered materials, such as insulating boron nitride [251-254] and high mobility graphene [255, 256], for studying new materials and physics. Moreover, 2H-TMDCs can be utilized to offer coupled spin-valley degree of freedom [257-261] and 2D ferromagnet is suitable for introducing magnetization [262-264]. Reversely, layered topological semimetals also bring unique building blocks to the 2D family, such as the protected surface states, chiral carriers of Weyl semimetals and strong SOC. Exotic physics may be expected in the future. For example, superconductivity in Weyl materials and QSHI may give rise to a new route for realizing non-abelian anyons, which is of great interest in quantum information science [265-267]. More importantly, it is clear that the exploration of the new physics harbored by layered topological materials and their heterostructures is still in its infancy and perhaps ahead of us is something far more exotic than what we have imagined.

## Acknowledgements

J. M. and D. S. thank the support by the National Natural Science Foundation of China (NSFC Grant Nos. 11674013,11704012) and the State Key Laboratory of Precision Measurement Technology and Instruments Fund for open topics. L. Z. and Z. L. acknowledge the support of the Singapore National Research Foundation under NRF award number NRF–RF2013-08 and MOE Tier 2 MOE2016-T2-2-153. S. W. acknowledges the support of the MIT Pappalardo Fellowship in Physics.